\documentclass[twocolumn]{aastex631}

\definecolor{darkblue}{rgb}{0.0,0.0,0.4}
\definecolor{darkgreen}{rgb}{0.0,0.4,0.0}
\hypersetup{linkcolor=black, citecolor=darkgreen, urlcolor=darkblue}

\received{March 28, 2024}
\revised{May 28, 2024}
\accepted{May 29, 2024}

\submitjournal{ApJ}

\usepackage{savesym}
\usepackage{soul}
\savesymbol{tablenum}
\usepackage{siunitx}
\usepackage{comment}
\restoresymbol{SIX}{tablenum}

\DeclareSIUnit \Msun {M_{\odot}}
\DeclareSIUnit \Lsun {L_{\odot}}
\DeclareSIUnit \LsunV {L_{\odot V}}
\DeclareSIUnit \pc {pc}
\DeclareSIUnit \kpc {kpc}
\DeclareSIUnit \mas {mas}
\DeclareSIUnit \yr {yr}
\DeclareSIUnit \pixel {pix}
\DeclareSIUnit \SN {SN}
\DeclareSIUnit \mag{mag}

\defcitealias{Nitschai_2023}{Paper I}
\defcitealias{Haeberle_2024}{Paper II}

\shorttitle{Metallicity analysis of $\omega$~Centauri}
\shortauthors{Nitschai et al.}

\begin{document}

\title{oMEGACat III. Multi-band photometry and metallicities reveal spatially well-mixed populations within $\omega$~Centauri's half-light radius.}

\correspondingauthor{M. S. Nitschai}
\email{nitschai@mpia.de}

\author[0000-0002-2941-4480]{M. S. Nitschai}
\affiliation{Max Planck Institute for Astronomy, K\"onigstuhl 17, D-69117 Heidelberg, Germany}
\author[0000-0002-6922-2598]{N. Neumayer}
\affiliation{Max Planck Institute for Astronomy, K\"onigstuhl 17, D-69117 Heidelberg, Germany}
\author[0000-0002-5844-4443]{M. H\"aberle}
\affiliation{Max Planck Institute for Astronomy, K\"onigstuhl 17, D-69117 Heidelberg, Germany}
\author[0009-0005-8057-0031]{C. Clontz}
\affiliation{Max Planck Institute for Astronomy, K\"onigstuhl 17, D-69117 Heidelberg, Germany}
\affiliation{Department of Physics and Astronomy, University of Utah, Salt Lake City, UT 84112, USA}
\author[0000-0003-0248-5470]{A. C. Seth}
\affiliation{Department of Physics and Astronomy, University of Utah, Salt Lake City, UT 84112, USA}
\author[0000-0001-7506-930X]{A. P. Milone}
\affiliation{Dipartimento di Fisica e Astronomia “Galileo Galilei,” Univ. di Padova, Vicolo dell’Osservatorio 3, Padova, I-35122, Italy}
\author[0000-0002-1212-2844]{M. Alfaro-Cuello}
\affiliation{Facultad de Ingenier\'{i}a y Arquitectura, Universidad Central de Chile, Av. Francisco de Aguirre 0405, La Serena, Coquimbo, Chile}
\author[0000-0003-3858-637X]{A. Bellini}
\affiliation{Space Telescope Science Institute, 3700 San Martin Drive, Baltimore, MD 21218, USA}
\author[0000-0001-6187-5941]{S. Dreizler}
\affiliation{Institut für Astrophysik und Geophysik, Georg-August-Universität Göttingen, Friedrich-Hund-Platz 1, 37077 Göttingen, Germany}
\author[0000-0002-0160-7221]{A. Feldmeier-Krause}
\affiliation{Max Planck Institute for Astronomy, K\"onigstuhl 17, D-69117 Heidelberg, Germany}
\affiliation{Department of Astrophysics, University of Vienna, T\"urkenschanzstrasse 17, 1180 Wien, Austria}
\author[0000-0003-2466-5077]{T.-O. Husser}
\affiliation{Institut für Astrophysik und Geophysik, Georg-August-Universität Göttingen, Friedrich-Hund-Platz 1, 37077 Göttingen, Germany}
\author{N. Kacharov}
\affiliation{Leibniz Institute for Astrophysics, An der Sternwarte 16, 14482 Potsdam, Germany}
\author[0000-0001-6604-0505]{S. Kamann}
\affiliation{Astrophysics Research Institute, Liverpool John Moores University, 146 Brownlow Hill, Liverpool L3 5RF, UK}
\author[0000-0002-7547-6180]{M. Latour}
\affiliation{Institut für Astrophysik und Geophysik, Georg-August-Universität Göttingen, Friedrich-Hund-Platz 1, 37077 Göttingen, Germany}

\author[0000-0001-9673-7397]{M. Libralato}
\affiliation{INAF, Osservatorio Astronomico di Padova, Vicolo dell’Osservatorio 5, Padova,I-35122, Italy}
\author[0000-0003-4546-7731]{G. van de Ven}
\affiliation{Department of Astrophysics, University of Vienna, T\"urkenschanzstrasse 17, 1180 Wien, Austria}
\author[0000-0001-6215-0950]{K. Voggel}
\affiliation{Universit\'{e} de Strasbourg, CNRS, Observatoire astronomique de Strasbourg, UMR 7550, F-67000 Strasbourg, France}
\author[0000-0003-2512-6892]{Z. Wang}
\affiliation{Department of Physics and Astronomy, University of Utah, Salt Lake City, UT 84112, USA}

\begin{abstract}

$\omega$~Centauri, the most massive globular cluster in the Milky Way, has long been suspected to be the stripped nucleus of a dwarf galaxy that fell into the Galaxy a long time ago. There is considerable evidence for this scenario including a large spread in metallicity and an unusually large number of distinct sub-populations seen in photometric studies. In this work, we use new MUSE spectroscopic and \textit{HST} photometric catalogs to investigate the underlying metallicity distributions as well as the spatial variations of the populations within the cluster up to its half-light radius. Based on 11,050 member stars, the [M/H] distribution has a median of $ (-1.614 \pm 0.003)$~dex and a large spread of $\sim$ 1.37~dex reaching from $ -0.67$~dex to $ -2.04$~dex for 99.7\% of the stars. In addition, we show the chromosome map of the cluster, which separates the red giant branch stars into different sub-populations, and analyze the sub-populations of the metal-poorest component. Finally, we do not find any metallicity gradient within the half-light radius, and the different sub-populations are well mixed.
\end{abstract}

\keywords{Galaxy nuclei(609) --- Globular star clusters(656) --- Star clusters(1567)}

\section{Introduction} \label{sec:intro}

\begin{figure*}[t!]
\plotone{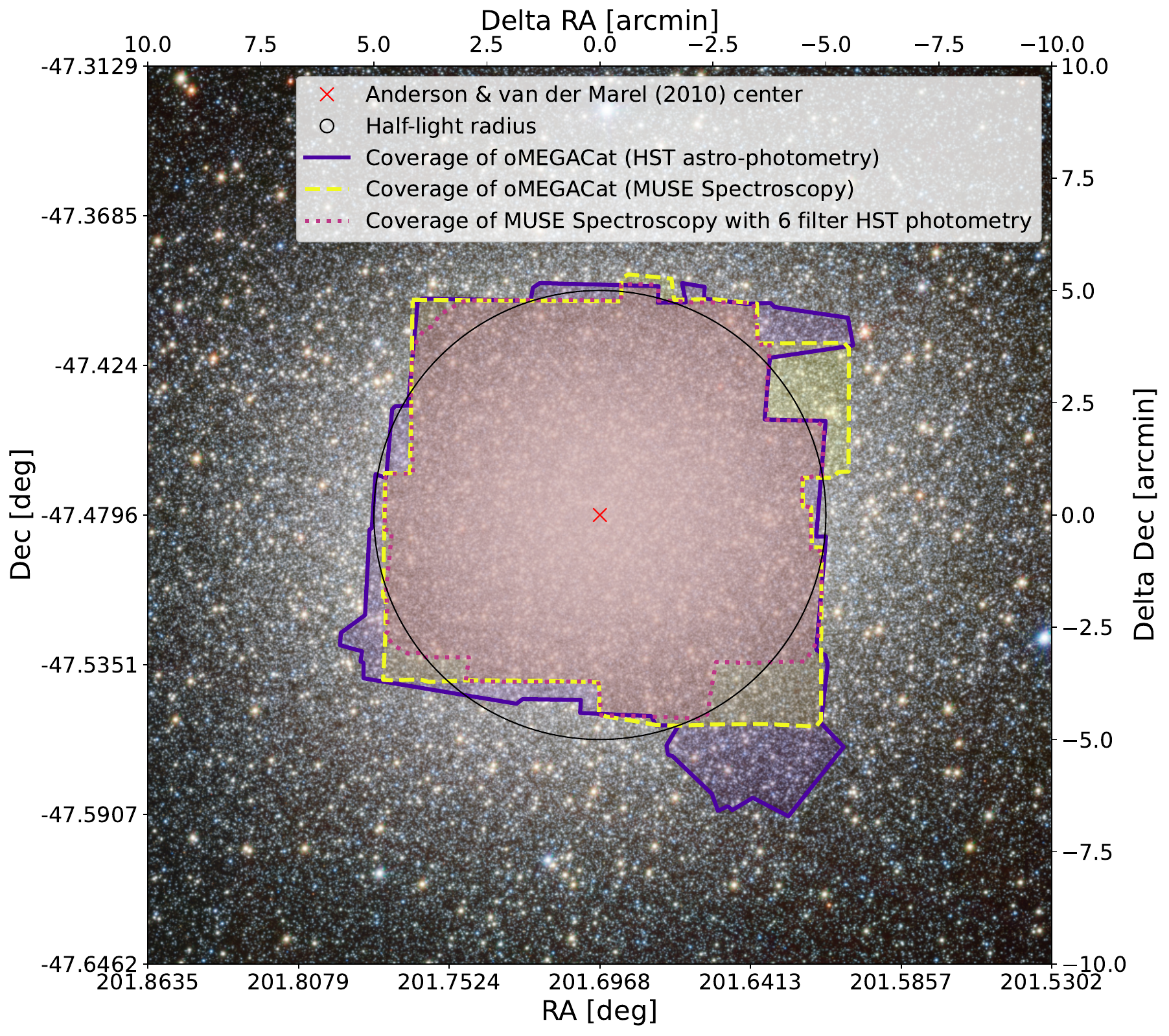}
\caption{\textbf{Footprint of data sets.} In dark purple with the solid line is the footprint of the new \textit{HST} catalog \citepalias{Haeberle_2024} and in yellow with a dashed line is the MUSE spectroscopic catalog \citepalias{Nitschai_2023} footprint. The light purple dotted contours show the area where we have a combined catalog with 6 filter photometry from \textit{HST}. The black circle is the half-light radius at 5~arcmin \citep{Harris_2010} and the red x mark is the \citet{Anderson2010} center. \textit{Background:} Cut-out of a wide-field image taken with the ESO/VST telescope (Image Credit: ESO/INAF-VST/OmegaCAM. Acknowledgement: A. Grado, L. Limatola/INAF-Capodimonte Observatory, \url{https://www.eso.org/public/images/eso1119b/})} \label{fig:footprints}
\end{figure*}

Nuclear star clusters (NSCs) are massive and compact star clusters in the innermost region of most galaxies \citep[see e.g. recent review by][]{Neumayer_2020}. They can be found in a wide range of galaxies, with an occupation fraction peaking at galaxy masses in the range of \SIrange{e8}{e10}{\Msun} \citep[e.g.][]{Sanchez-Janssen_2019, Hoyer_2021}. Due to their small sizes (half-light radii of \SIrange{1}{10}{\pc}) and sizable masses (\SIrange{e6}{e8}{\Msun}, \citealt{Georgiev_2014, Neumayer_2020}), they are the densest stellar systems in the Universe, $\geq$ \SI{e6}{\Msun\per\cubic\pc} \citep{Walcher_2005, Neumayer_2020}. Moreover, they have extended star formation histories \citep{Walcher_2006, Seth_2006, Kacharov_2018}.

Stripped nuclei are created when smaller galaxies fall into a larger one and are disrupted by tidal forces. During this process they lose most of their stellar content, however, similar to globular clusters, NSCs survive these mergers because they are highly compact and dense. These stripped nuclei will look like massive globular clusters, and hence be able to hide amongst the globular clusters in a galaxy. Semi-analytic models predict 2 to 6 stripped nuclei in our Milky Way halo \citep{Pfeffer_2014, Kruijssen_2019}. One example is M~54, the nucleus of the Sagittarius dwarf galaxy \citep{Alfaro-Cuello_2019, Alfaro-Cuello_2020, Kacharov2022}; it was first discovered as a globular cluster until the detection of the remains of the galaxy \citep{Ibata_1994}, which is undergoing tidal stripping over the last several \si{\giga\yr} \citep{Ibata_1997, Laporte2018}.

The most promising other candidate for a stripped nucleus in the Milky Way halo is $\omega$~Centauri ($\omega$~Cen, NGC~5139), due to the complexity of its stellar populations. It shows multiple sequences in its color-magnitude diagram \citep[e.g.][]{Anderson1997, Bedin2004, Bellini_2010, Milone_2017} and has a large spread in metallicity \citep[e.g.][]{Freeman1975, Johnson_2010}. In addition, the internal kinematics of the cluster show the presence of a central stellar disk and a bias toward tangential orbits in the outer parts \citep{VanDeVen_2006}, counter rotation in the very central region \citep{Pechetti_2024}, and fast-moving stars in the inner 3 arcsec \citep{Haeberle_2024fs} indicating the presence of an intermediate-mass black hole. In addition, the orbit of $\omega$~Cen has been associated with the \textit{Gaia}-Enceladus merger \citep{Massari_2019, Pfeffer_2021, Callingham_2022, Limberg2022} $\sim$ 10~Gyr ago \citep{Helmi2018}. The star formation and assembly history of the cluster is still debated and the claims for its populations age span vary from $2-3$ Gyr \citep{Hilker_2004} up to $4-5$ Gyr \citep{Villanova_2007}. Other studies have found models fully consistent with a shorter star formation of \SIrange{1}{2}{\giga\yr} \citep{Joo_2013}. Finally, tidal material associated with $\omega$~Cen is directly seen \citep[e.g.][]{majewski_2012, Ibata_2019, Ibata_2023} which further strengthens the stripped nucleus scenario.

$\omega$~Cen is the brightest, most massive globular cluster in the Milky Way \citep[$\sim$ \SI{3.55e6}{\Msun},][]{Baumgardt2018}. Since it is only at a distance of $\sim$ \SI{5.43}{\kpc} \citep{Baumgardt2021}, and not heavily obscured by dust \citep[$E(B-V)=0.12$]{Schafly_2011}, it provides us with an ideal laboratory to study a NSC. Further, because the masses and star formation histories of NSCs track the galaxies they lived in \citep[e.g.][]{Kacharov_2018, Sanchez-Janssen_2019} and are much longer lived than stellar streams, it can also provide us with valuable information about the host galaxy and its merger with the Milky Way.

\begin{figure*}[t]
\plottwo{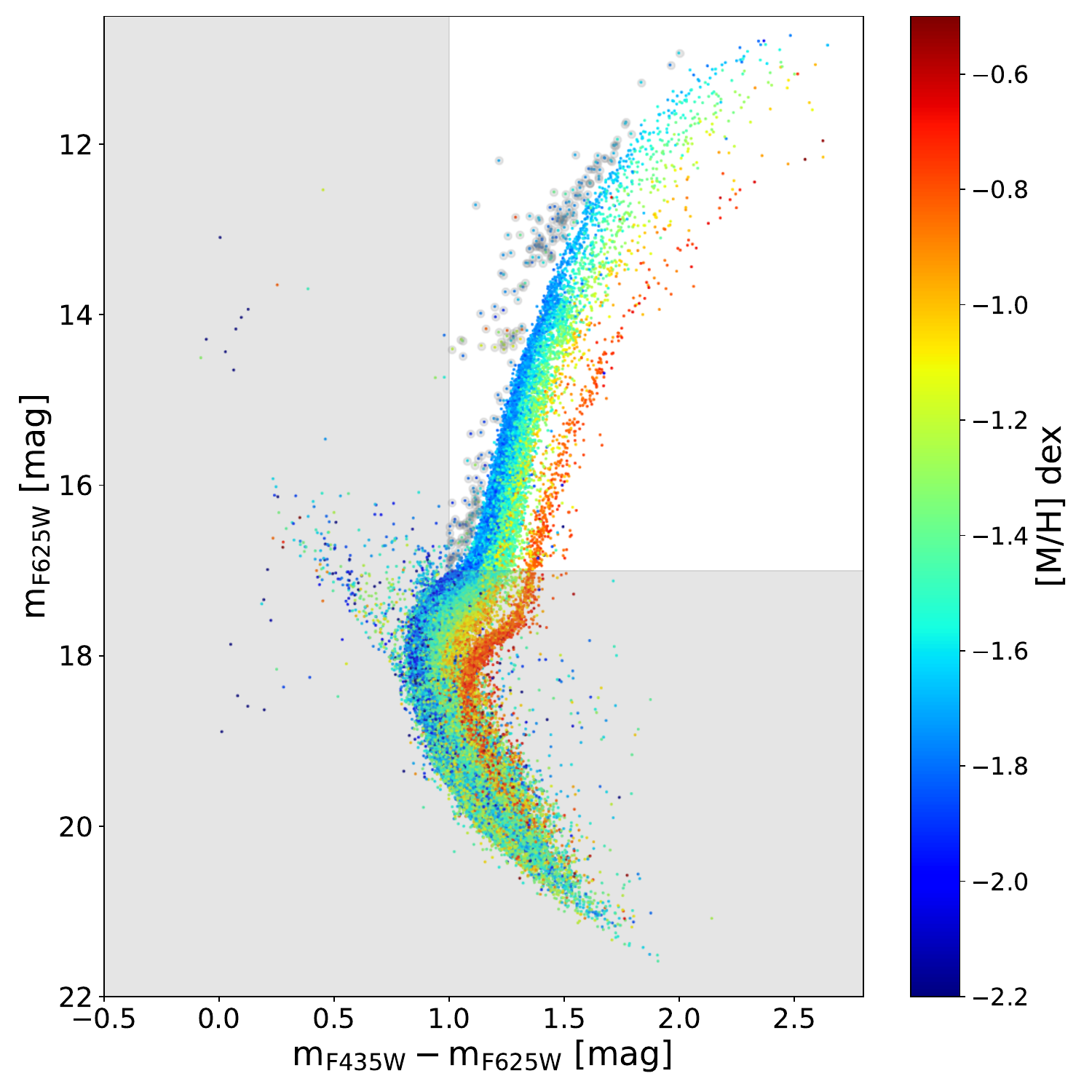}{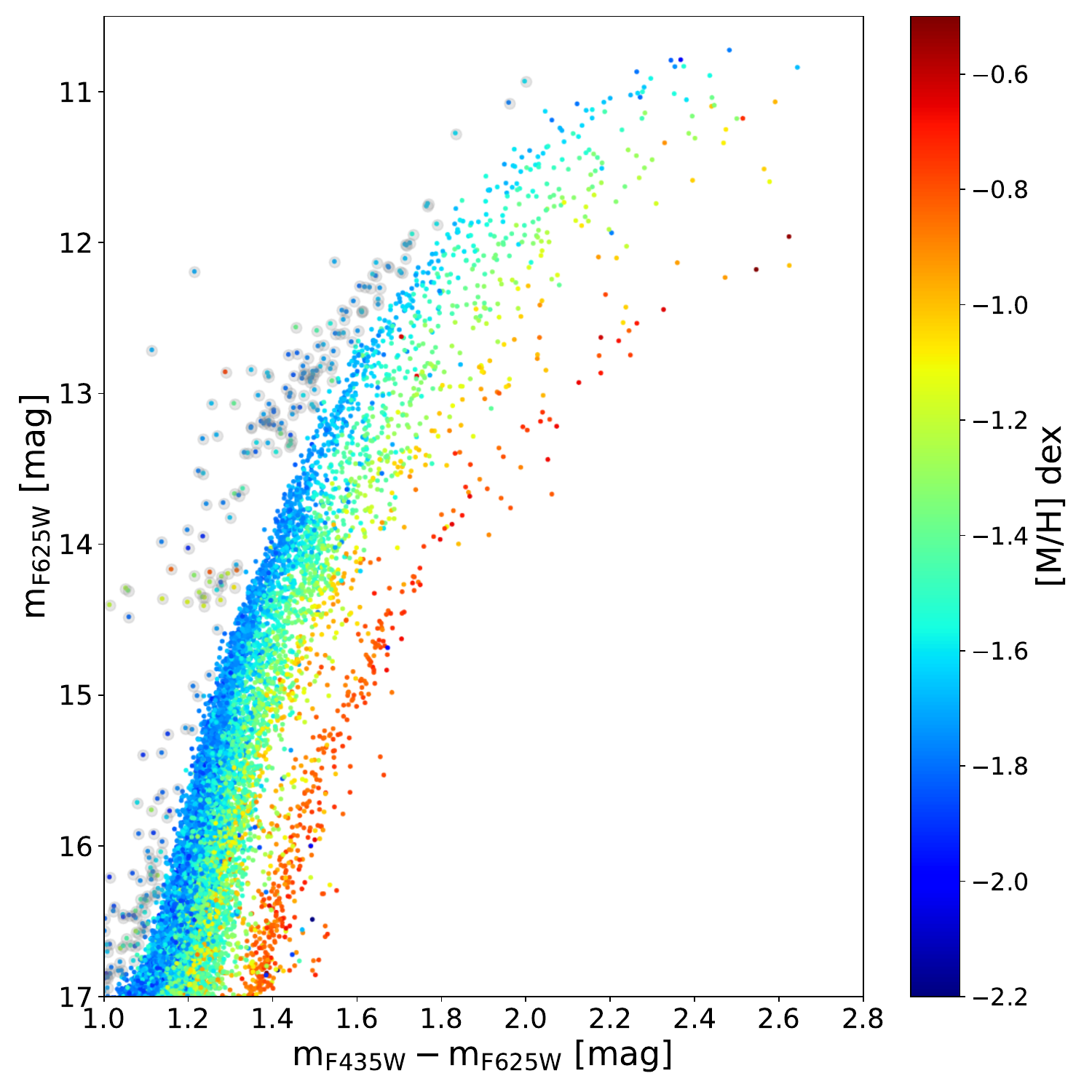}
\caption{\textbf{Color-magnitude Diagrams (CMDs)}. On the left is the full CMD of our combined data set color-coded with the metallicity (with 145,531 stars). On the right side is a zoom-in on the red giant branch. The gray shaded areas indicate the regions excluded in this work and gray circle dots on the left of the RGB are the excluded AGB and (evolved) BSSs from the RGB sample. \label{fig:CMD}}
\end{figure*}

In our oMEGACat project, we study $\omega$ Cen in great detail with the aim to explore its formation history and interactions with the Galaxy. To do that, we created a spectroscopic catalog \citep[][from here on Paper I]{Nitschai_2023} using the Multi-Unit Spectroscopic Explorer (MUSE), and a photometric and astrometric catalog \citep[][from here on Paper II]{Haeberle_2024} using the \textit{Hubble Space Telescope (HST)} up to the half-light radius of the cluster (see \autoref{fig:footprints} for the coverage of each catalog). In this work, we make use of these data and investigate the metallicity information we have for the stars in $\omega$~Cen. In \autoref{sec:data} we summarize the observations used to create the catalogs and the quality cuts used in this specific work. Afterwards in \autoref{sec:results}, we present the full metallicity distribution, the chromosome map and we investigate any possible spatial variation of the metallicity. Finally, in \autoref{sec: concl} we summarize and discuss our findings.

\section{Data} \label{sec:data}

\subsection{Spectroscopic data}\label{sec: spec}

A detailed description of the spectroscopic data used can be found in \citetalias{Nitschai_2023}. In summary, the spectroscopic data were acquired with MUSE \citep{Bacon2010, Bacon2014}, a second-generation Very Large Telescope (VLT) instrument mounted on the UT4 at the Paranal Observatory in Chile observing in the optical domain (\SI{480}{\nano\meter}-- \SI{930}{\nano\meter}). The observations were carried out between February 2021 and September 2022 in run 105.20CG.001 (PI: N. Neumayer) and in addition, we also used the complementary MUSE guaranteed time observations (hereafter “GTO data”). These data together have a full coverage out to the half-light radius \citep[4.65' or \SI{7.04}{\pc},][]{Baumgardt2018} of the cluster.

For the analysis of the data we first extracted the spectra for the individual stars using \textsc{PampelMuse}\footnote{\url{https://pampelmuse.readthedocs.io/en/latest/about.html}} \citep{Kamann2013} and the \textit{HST} catalog from \citet{Anderson2010}. Afterward, we used \textsc{spexxy}\footnote{\url{https://github.com/thusser/spexxy}} \citep{Husser_2016} to measure the physical parameters of the stars such as effective temperature, metallicity, and line-of-sight velocity. The observed spectra are compared to synthetic spectra from the Phoenix library \citep{Husser_2013} with varying log(g), $\rm T_{\rm eff}$, [M/H] while minimizing the $\chi^2$ difference. During the fit with \textsc{spexxy}, log(g) is fixed to the value provided by the isochrone from the PARSEC database \citep{Marigo_2017} used to estimate the initial guesses, the $\alpha$-enhancement is kept constant at [$\alpha$/Fe]=0.3\,dex, while $T_{eff}$, [Fe/H] and the line-of-sight velocity are determined.

In addition, we performed multiple tests to check for the robustness of our results, which included an error analysis and membership determination, as well as necessary corrections, like the atomic diffusion correction (ADC). For more details and the spectroscopic data themself, we refer to \citetalias{Nitschai_2023}.

\subsection{Photometric data}\label{sec: phot}

The detailed description of the photometric data is available in \citetalias{Haeberle_2024} and here we only briefly summarize it. The astro-photometric part of the catalog is based on observations with the \textit{HST} spanning over 20 years and using the Advanced Camera for Surveys Wide Field Channel (ACS/WFC), and the Wide Field Camera 3 UVIS Channel (WFC3/UVIS). The data was obtained for various general observing and calibration programs, including a new dedicated program (GO-16777, PI: A. Seth) aimed at providing complementary filter coverage out to the half-light radius and additional epochs required for proper motion measurements. All {\it HST} data used for the creation of the astro-photometric catalog can be found under the following DOI in the Mikulski Archive for Space Telescopes (MAST): \dataset[10.17909/26qj-g090]{http://dx.doi.org/10.17909/26qj-g090}.

The catalog contains both high-precision photometry spanning from the UV to the near-infrared and proper motions for around 1.4~$\times~10^6$ stars. The photometric measurements were performed using techniques specifically designed to analyze crowded environments like the cores of GCs (see \citealt{2017ApJ...842....6B} for details). Six filters have coverage over the full field (ACS/WFC: F435W, F625W, F658N; WFC3/UVIS: F275W; F336W, F814W), in addition, the central region is also covered with multiple epochs of the WFC3/UVIS F606W filter. The proper motions were determined using the technique established and improved in \citet{2014ApJ...797..115B, Bellini_2018, 2018ApJ...861...99L} and \citet{2022ApJ...934..150L}.

In \citetalias{Haeberle_2024} an empirical correction for spatially dependent photometry variations has been calculated for all seven filters, which is needed due to differential reddening and systematic zero-point variations caused by instrumental effects.
 Whenever we use this new \textit{HST} photometry we add the correction for differential reddening and instrumental effects to the magnitudes except where explicitly noted. The coverage for the combined data catalog with 6-filter \textit{HST} photometry is shown in \autoref{fig:footprints}.

To combine the photometry from \citetalias{Haeberle_2024} with the [M/H] values from \citetalias{Nitschai_2023}, we perform an astrometric crossmatch between the two catalogs using a matching radius of 40\,mas (equivalent to approximately one WFC3/UVIS pixel). In addition, we require a mag difference in the F625W and F435W filters of $\Delta$mag$<$0.1 between the new photometry and the \cite{Anderson2010} photometry used for the extraction of the spectra. This leaves us with 307,030 matched stars out of the total of 342,797 from the catalog published in \citetalias{Nitschai_2023}.

\subsection{Quality cuts}\label{sec: quality}

As default, we assume the quality cuts for the spectroscopic catalog described in \citetalias{Nitschai_2023}, which are given as an extra flag in the catalog (`Flag'). In summary, we use stars with a minimum SNR of 10, $>$95\% membership probability, and with a relative accuracy of recovered magnitude from spectrum extraction with \textsc{PampelMuse} $\geq 0.6$. We also exclude stars near the edge of the field ($<$ 5 pixel) and where the cross-correlation for the velocity is not reliable. These cuts decrease the number of stars from 342,797 to 156,871. The color-magnitude diagram (CMD) in the left panel of \autoref{fig:CMD} shows 145,531 stars that fulfill these criteria and are matched to the \textit{HST} photometry as described in the previous section.

As discussed in \citetalias{Nitschai_2023} the above cuts are a compromise that allow us to keep a high number of stars with reliable measurements. Since we want to study the metallicity distribution, chromosome map, and spatial variations of sub-populations we want precise and unbiased [M/H] values over our whole field of view. Hence in addition to the default cuts, we use only stars brighter than a general magnitude cut at $m_{\rm F625W} \le 17$. This restriction to bright stars aids in avoiding completeness issues of the spectroscopic data set at fainter magnitudes (see \autoref{ap:compl}), getting higher precision on the [M/H] values with a median SNR of $\sim$~54, and removes any [M/H] biases between the different original data sets (GTO, GO, NFM, see \autoref{ap:bias}), that can be caused by different exposure times. To obtain a clean red giant branch (RGB) sample, in a further step, we remove by eye (evolved) blue straggler stars (BSSs) or asymptotic giant branch (AGB) stars. This selection restricts the sample to 11,050 RGB stars (see the right panel in \autoref{fig:CMD}).

We do not expect a significant fraction of binaries within our sample. In general, $\omega$~Cen has a low binary fraction, only 5\% \citep{Elson_1995, Mayor_1996} and even lower in recent work at 2.70\% $\pm$ 0.08\% \citep{Bellini_2017b} and 2.1 $\pm$ 0.4 \% \citep{Wragg_2024}. Therefore, we can safely assume that binaries will not have a significant effect on our results, especially since \citet{Wragg_2024} find similar binary fractions for all stellar evolutionary stages using the GTO subsample of the oMEGACat catalog \citep{Nitschai_2023}. Only the BSSs show an enhanced binary fraction ($>20\% $). Because of the higher binary fraction among BSSs and the chemically peculiar atmospheres of hot stars due to diffusion 
\citep[at $T_{\rm eff} \gtrsim$ 7,800~K, ][]{Lovisi_2013}, we exclude all stars bluer than the main RGB track (see gray areas in \autoref{fig:CMD}.)

\begin{deluxetable}{lcc}
\tabletypesize{\scriptsize}
\tablewidth{0.80\columnwidth}
\tablenum{1}
\tablecaption{Data samples \label{tab:data}}
\tablehead{
\colhead{Name} & \colhead{Number of stars} & \colhead{Description}}
\startdata 
MUSE & 342,797 &  spectroscopy \citetalias{Nitschai_2023}\\
HST matched &  307,030 &  spectroscopy with \\ & &  astro-photometry \citetalias{Haeberle_2024} \\
MUSE QC & 156,871 &  QC of spectroscopy\\
HST QC &  145,531 & HST matched with \\ &&QC MUSE \\
 RGB &  11,050&  HST QC and $m_{\rm F625W} \le 17$ \\
 RGB phot &  10,850 &  RGB that has a measurement in\\ &&  F625W, F435W, F275W, F336W \\ &&  and $m_{\rm F814W} \le 17$\\
\enddata
\tablecomments{QC: quality cuts described in \citetalias{Nitschai_2023} and beginning of \autoref{sec: quality}.  The RGB phot sample is needed to have RGB stars in all filters required for the Chromosome map, see \autoref{sec:chromosome}. }
\end{deluxetable}

With these cuts, we get a mean [M/H] error of 0.041~dex. Further, by comparing our MUSE data to the photometric catalog used for the extraction of the spectra \citep{Anderson2010} we find a completeness $>$80\% in all individual cubes, with a median completeness at 17~mag of 97\%. We spatially bin the sample stars and find no significant variation in the mean magnitude with radius or at any point within the field of view. This suggests we can make spatial comparisons with minimal concern about instrumental issues.

Note that all metallicities unless otherwise indicated include the ADC correction, we always use the scaled errors for the metallicity as described in \citetalias{Nitschai_2023} Section 4.2 and the new photometry from \citetalias{Haeberle_2024}.

In the following section, we will investigate the metallicity information we have for the cluster in detail using the above quality cuts. All the data samples described in this section are summarized in \autoref{tab:data}, in \autoref{sec:results} we will use either the RGB or RGB phot sample.

\begin{figure*}[t]
\plottwo{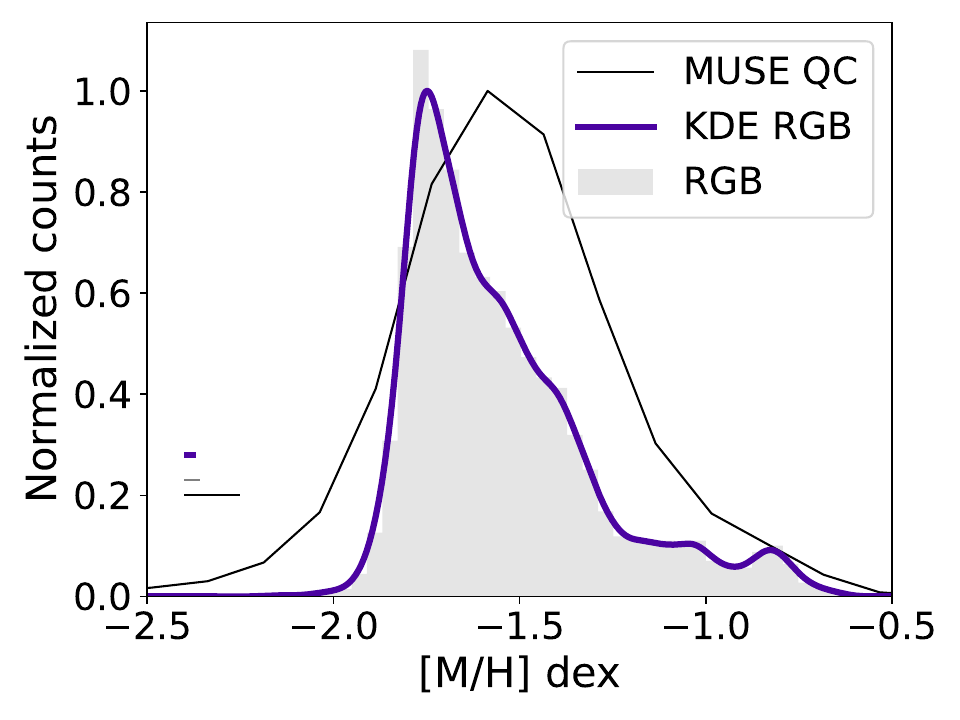}{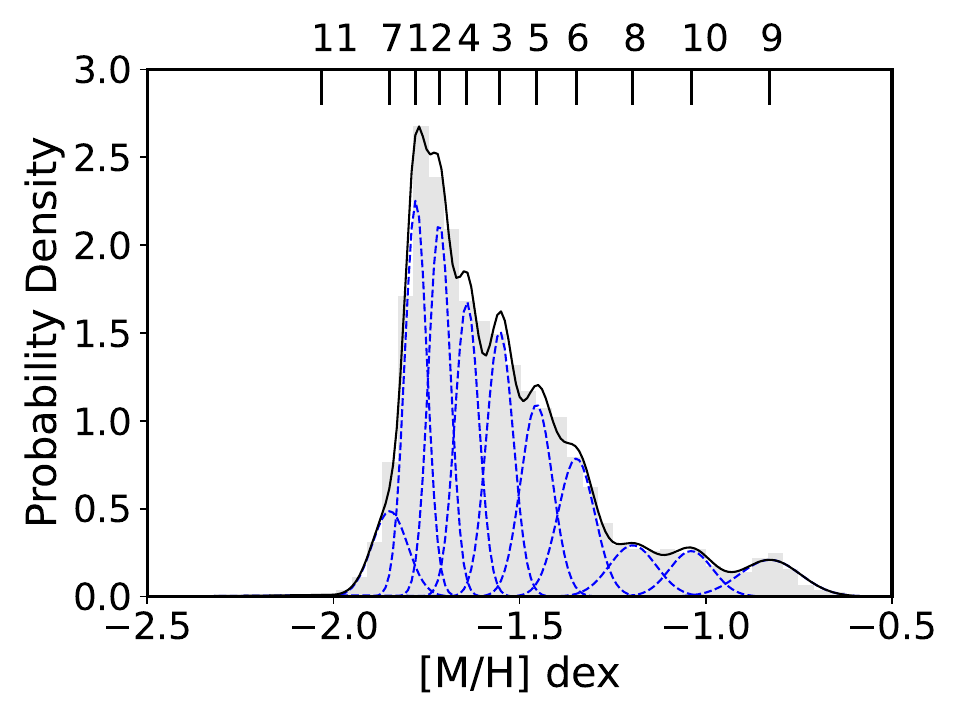}
\caption{\textbf{Metallicity distribution}. The figure on the left shows the normalized distribution of [M/H] values in the spectroscopic catalog \citepalias{Nitschai_2023} using the default quality cuts in the black line (MUSE QC), using the extra condition of a minimum brightness of 17~mag in the gray shaded histogram (RGB) and the purple line is the KDE estimator for that distribution. The small horizontal lines on the left side show the bandwidth, 0.030~dex, for the distributions, which are close to the mean errors of the [M/H] values. On the right, we show the same distribution (RGB) as probability density but fitted with a Gaussian mixture model, shown as a black solid line, with each component (11 in total) shown as a blue dashed line. Note that one component has a small amplitude and is not visible in this plot.\label{fig:mh distr}}
\end{figure*}

\section{Results}\label{sec:results}

\subsection{Metallicity distribution} \label{sec:distribution}

First, we analyze the total metallicity distribution of the 156,871 stars (MUSE QC) passing the default criteria, see the left side of \autoref{fig:mh distr}. The additional magnitude-cutoff ($m_{\rm F625W}\leq$17~mag) leaves us with 11,050 stars (RGB) and decreases the more extended tail to lower values (below $\sim -2$ dex) but also shifts the peak of the distribution to lower values since on average the metal-rich stars have a main-sequence turn-off at fainter F625W magnitudes and hence more metal-rich stars are removed. The cut however ensures that both GTO and GO data sets have a similar distribution and no bias is caused due to different data sets, see \autoref{ap:bias}. We also used a Kernel Density Estimator (KDE) to estimate the probability density of the distribution, which we normalized to one to plot it onto the same figure. In this distribution multiple peaks are visible and we will identify them later in this section.

The mean [M/H] value is at ($ -1.550 \pm 0.002$)~dex and the intrinsic standard deviation is ($ 0.255\pm 0.002$)~dex, which is close to the mean value of the distribution with no magnitude cut ($-1.53$~dex). The median is ($ -1.614 \pm 0.003$)~dex with half of the differences between the 16th and 84th percentile being ($ 0.221 \pm 0.003$)~dex. The median of the overall distribution without a magnitude cut is $-1.54$~dex, which is closer to the mean value since more metal-rich stars are included. All the values and errors (68\%) were calculated with bootstrapping.

The range in metallicity for stars brighter than 17~mag reaches from $\sim$ $-2$~dex to almost $-0.5$~dex. In detail, 68\% of the stars are between $ -1.77$~dex and $ -1.33$~dex and 99.7\% between $ -2.04$~dex and $ -0.67$~dex. This huge metallicity spread is, as mentioned previously, an indication of multiple stellar populations, which indeed are visible as distinct sequences in the color-magnitude diagram (CMD, see \autoref{fig:CMD}).

All our results are comparable to the values in \citet{Husser_2020} where in a smaller sample of 1,247 stars they found a mean value of $-1.50$~dex, a median of $-1.65$~dex and a range from $-1.82$ to $-1.31$~dex for the 1st and 3rd quartiles. However, since our sample is 10 times larger we have even more accurate values with small uncertainties.

We also used a Gaussian Mixture Model (GMM) to describe our distributions and the result can be seen on the right side of \autoref{fig:mh distr} and in \autoref{tab:multiGaus}. We find that the best number of Gaussians to describe the distribution is eleven using the Bayes Information Criterion (BIC), but eight components would give an almost equally good fit. However, one component is not visible in the Figure (index 11 in the Table) since it has a very small amplitude compared to the rest and would not count as a separate population but is needed to describe the distribution in the model. Similarly, component 7 is needed for the extended tail towards lower metallicities but together with components 1, 2, and 4 they are not distinguishable by eye in the highest peak of the distribution. Components 3, 8, 9, and 10 can be seen by eye as smaller peaks, and components 5, and 6 are more difficult to distinguish but are still visible.

\begin{deluxetable}{lccc}
\tablewidth{2.0\columnwidth}
\tablenum{2}
\tablecaption{Multi-Gaussian Components of the metallicity distribution \label{tab:multiGaus}}
\tablehead{
\colhead{index} & \colhead{Mean} & \colhead{Intrinsic standard deviation} & \colhead{Fraction of stars}\\
\colhead{\#} & \colhead{[dex]} & \colhead{[dex]} & \colhead{\%}}
\startdata 
1 &  -1.779 &  0.030 & 18.2\\ 
2 &  -1.715 &  0.030 &  16.0\\
3 &  -1.553 &  0.036 &  13.8\\
4 &  -1.642 & 0.032 &  13.4\\
5 &  -1.454 & 0.043 &  11.8\\
6 &  -1.348 &  0.049 &  9.7\\
7 &  -1.849 &  0.048 &  4.6\\
8 &  -1.196 &  0.063 &  4.3\\
9 &  -0.828 & 0.082 &  4.2\\
10 &  -1.039 &  0.058 &  3.8\\
11 &  -2.033 &  0.210 &  0.3\\
\enddata
\tablecomments{The index \# is assigned from the highest to the lowest fraction of stars.}
\end{deluxetable}

We have compared our metallicity findings with previous studies already in \citetalias{Nitschai_2023} where we made a one-to-one comparison. Also, the overall distribution agrees with previous studies \citep[e.g.][]{Johnson_2010, Meszaros_2021, Alvarez_Garay_2024}, even though we find more peaks in our distribution. \citet{Johnson_2010} find 5 peaks in the distribution at $-1.75, -1.50, -1.15, -1.05$ and $-0.75$ dex, while \citet{Meszaros_2021} and \citet{Alvarez_Garay_2024} find four peaks, $-1.65, -1.35, -1.05$ and $-0.7$ dex, and $-1.85, -1.55, -1.15$, and $-0.80$ dex respectively. All these peaks are in agreement with at least one of our peaks within the 3 $\sigma$ range. The difference in the number of peaks can be explained due to the difference in the data, specifically the amount of data, since it can make it difficult to identify and separate peaks in the distribution with a lower amount of stars. All the above-mentioned studies have stars that reach larger radii than ours, but have fewer than 1,000 stars, while we have more than 10 times their number which allows us to distinguish different metallicity components more easily. However, even in our case, one could reduce the number of components to five if one only looks at the distribution by eye, since peaks 1, 2, 4, and 7 can be taken as one, and components 5 and 6 or 8 and 10 are not that clearly visible peaks in the overall distribution and component 11 is not at all visible. Even with the GMM, we can see that eight components would give an almost as perfect fit to the distribution and that small variations in the sample of stars can also change the preferred number of Gaussians. This demonstrates the advantage of our huge data set that allows for much more detailed studies and analysis.

\begin{figure*}[t]
\plotone{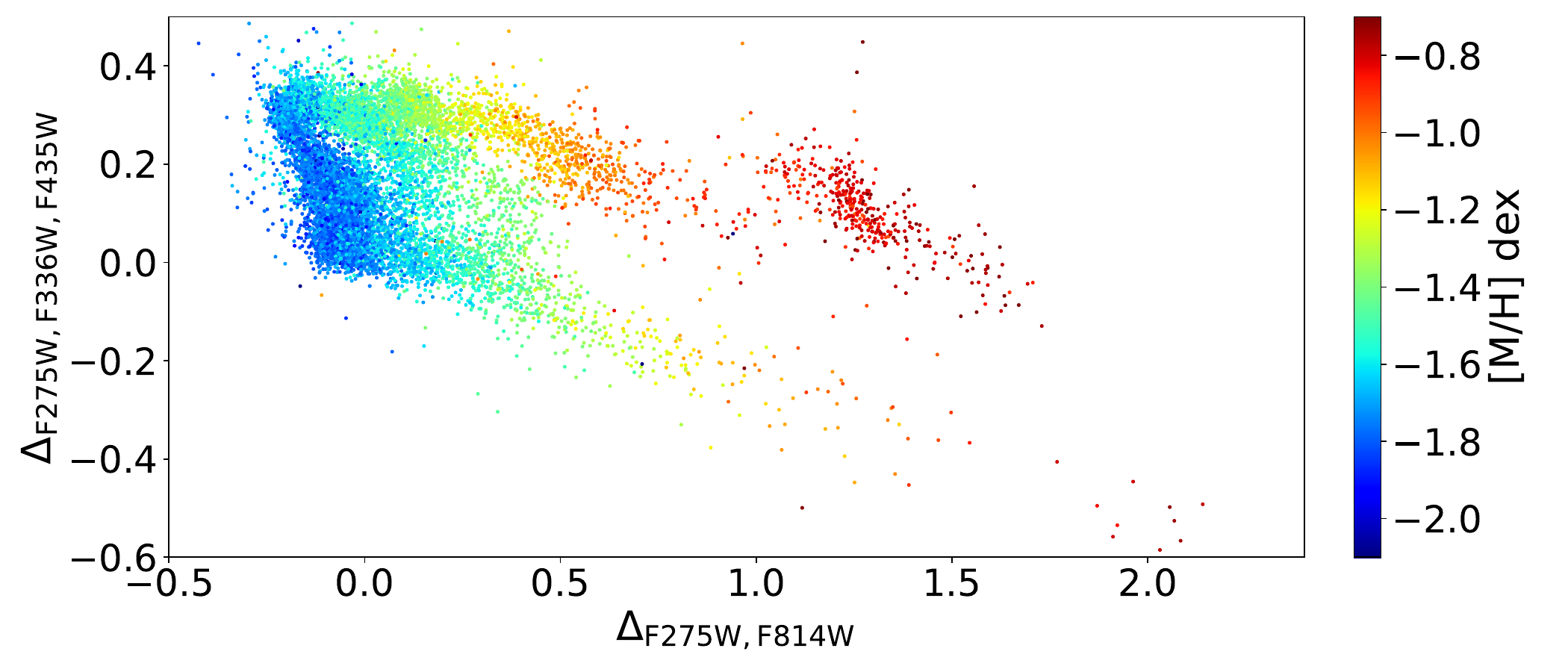}
\caption{\textbf{Chromosome map of $\omega$~Cen}. The chromosome map of $\omega$~Cen is color-coded with metallicity. One can identify different sub-populations in the map but since we do not aim to separate them precisely in this work we also do not draw the line separating first and second-generation stars as it is usually done in globular clusters. \label{fig:chr}}
\end{figure*}

In \citet{Husser_2020} they identify nine sub-populations and give their mean metallicity values. Our multi-Gaussian Mixture Model does not separate the stars in different sub-populations, since also photometric analysis needs to be taken into account. However, we can compare their position on the chromosome map in \autoref{sec:chromosome} and identify which component belongs to which sub-population (P) identified in \citet{Husser_2020}. P1, P2 and P3 (mean [Fe/H]= $-1.83, -1.80, -1.74$~dex) would correspond to our components 7, 1 and 2, P5 ($-1.24$~dex) to component 6, P8($-1.50$~dex), P6($-1.47$~dex) and P4($-1.50$~dex) to components 3 and 5. Our component 6 could be part of P4 or P3. From the chromosome map, we can identify component 9 as P7 ($-0.18$~dex) and components 10 and 8 as P9 ($-0.72$~dex) but the metallicity values disagree the most with our findings. However, they have a lot fewer stars than we have especially since their P7 and P9 are not well populated. With our data, they can be much better studied and separated. In the following sections, we investigate a bit further different sub-populatins.  We do not aim to precisely separate and identify all sub-populations in detail, this will be done in future work (Clontz et al., in prep). Nevertheless, this analysis already shows how our huge data set can improve sub-population studies.

Lastly, in \citet{Johnson_2020} they find two very metal-poor populations, below $-2$ and $-2.25$~dex. Our component 11 contains 30 stars at the very low end of the metallicity distribution reaching as low as $-2.738$~dex and as high as -1.990~dex but only 4 stars above $-2.25$~dex. However, this component is just a tiny fraction of the whole sample and its stars do not cover the whole CMD range (most of them lie close to magnitude 17~mag), indicating that they are not a separate population and are not the same stars as in \citet{Johnson_2020} where most of these stars lie on the brighter RGB.
Hence, our component 11 stars are individual stars with special [M/H] values either due to measurement problems or they could be binary stars. Overall, we can conclude that we do not find the most metal-poor population from \citet{Johnson_2020} in our sample, even though we have a few low metallicity stars.

\subsection{Photometry and Metallicity}\label{sec:phot_metal}
\subsubsection{Chromosome map}\label{sec:chromosome}

In this subsection, we focus again on RGB stars with $m_{\rm F625W}<17$\,mag. In addition, we require the stars to have accurate photometric measurements in  5 different \textit{HST} filters (F275W, F336W, F435W, F625W, and F814W) in the photometric catalog of \citetalias{Haeberle_2024}. These requirements leave us with a sample of 10,850 stars (RGB phot).

Almost all globular clusters host significant light-element abundance spreads within them, following a basic pattern, a population with abundances akin to field stars, and one or more anomalous populations, with varying enrichment in some elements (e.g., He, N, Na) and depletion in others (e.g., C, O), but the specifics of each cluster are unique \citep[see the review by][]{Bastian_2018}. These abundance spreads are typical for old and massive clusters and are known as multiple populations. However, abundance variations in Fe and heavy elements are rare and only have been detected in more complex clusters such as $\omega$\,Cen \citep{Willman_2012}.

The chromosome map is a photometric diagram used to identify multiple stellar populations in globular clusters. It uses multi-color photometric information that is sensitive to element abundance variations to characterize the presence and diversity of multiple populations \citep{Milone_2017b}. Many studies have provided accurate investigations of the sub-populations of $\omega$\,Cen \citep[e.g.][]{Tailo_2016, Latour_2021, Husser_2020}. Here, we take advantage of an unprecedented dataset that comprises photometry and spectroscopy of more than 10,000 RGB stars to better investigate them. 

To construct the chromosome map, we follow the technique from \citet{Milone_2017} and use the filters F814W, F336W, F275W, and F435W. In \citet{Milone_2015a, Milone_2015b} it has been shown that the combination of the pseudo-color $C_{\rm F275W, F336W, F438W}$ (in our case $C_{\rm F275W, F336W, F435W}$) \citep{Piotto_2015} with the $m_{\rm F275W} - m_{\rm F814W}$ color maximizes the separation between stellar populations along the main sequence and the RGB. Here we analyze only the RGB populations, deferring a more comprehensive view of sub-populations and their exact separation in the cluster to a later paper (Clontz et al., in prep). 

In detail, the pseudo-color $C_{\rm F275W, F336W, F435W}$ is selected since it is sensitive to the degree of CNO processing in the multiple populations. This is thanks to the fact that the F275W filter includes the OH molecular band, the F336W the NH band, and the F435W the CH and CN bands \citep[e.g.,][]{Milone_2015a}. Simultaneously the F275W and F814W filters provide a wide color range being sensitive to the effective temperature and thus to different metallicity or helium abundances, since helium enhanced stars are hotter at the same luminosity \citep[e.g.,][]{Milone_2015a, Milone_2012}.


The detailed steps to calculate the $\Delta$ values can be found in \autoref{sec: detail chr}. 
In general, the following relations are used:
\begin{eqnarray}
    \Delta_{\rm F275W,F814W} = W_{\rm F275W, F814W} \cdot \nonumber\\ \frac{X-X_{fiducialR}}{X_{fiducialR}-X_{fiducialB}}
\end{eqnarray}
and
\begin{eqnarray}
    \Delta_{\rm F275W,F336W, F435W} = W_{\rm F275W,F336W, F435W} \cdot \nonumber\\ \frac{Y_{fiducialR}-Y}{Y_{fiducialR}-Y_{fiducialB}}
\end{eqnarray}
with fiducialR the redder fiducial line and fiducialB the bluer line in color space. Further, $X = m_{\rm F275W} - m_{\rm F814W}$, $Y = C_{\rm F275W, F336W, F435W}$ and the pseudo-color $C_{\rm F275W, F336W, F435W} =(m_{\rm F275W} - m_{\rm F336W}) - (m_{\rm F336W} -m_{\rm F435W})$. The final chromosome map is shown in \autoref{fig:chr}. The few stars lying on the very blue side of the chromosome map are most likely remaining evolved BSSs or/and binaries \citep[see][]{Marino_2019NGC3201,Martins2020, Kamann_2020} that have not been removed from our RGB sample because they lie close to the bluest RGB stars. Including the bright (evolved) BSSs (gray shaded points in \autoref{fig:CMD}) would create a more extended tail to bluer $\Delta_{\rm F275W,F814W}$ values in the chromosome map.

\begin{figure*}[t]
\plottwo{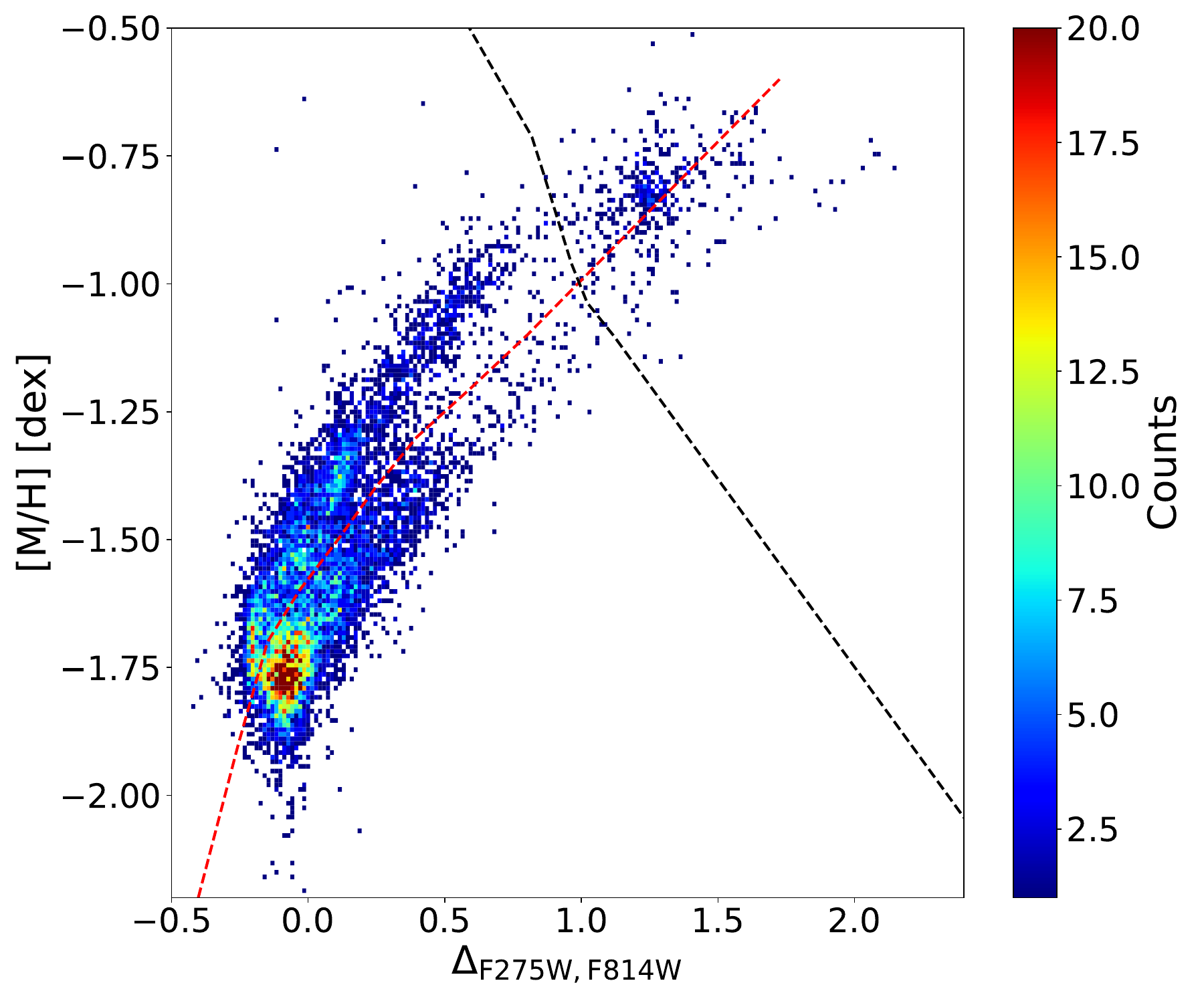}{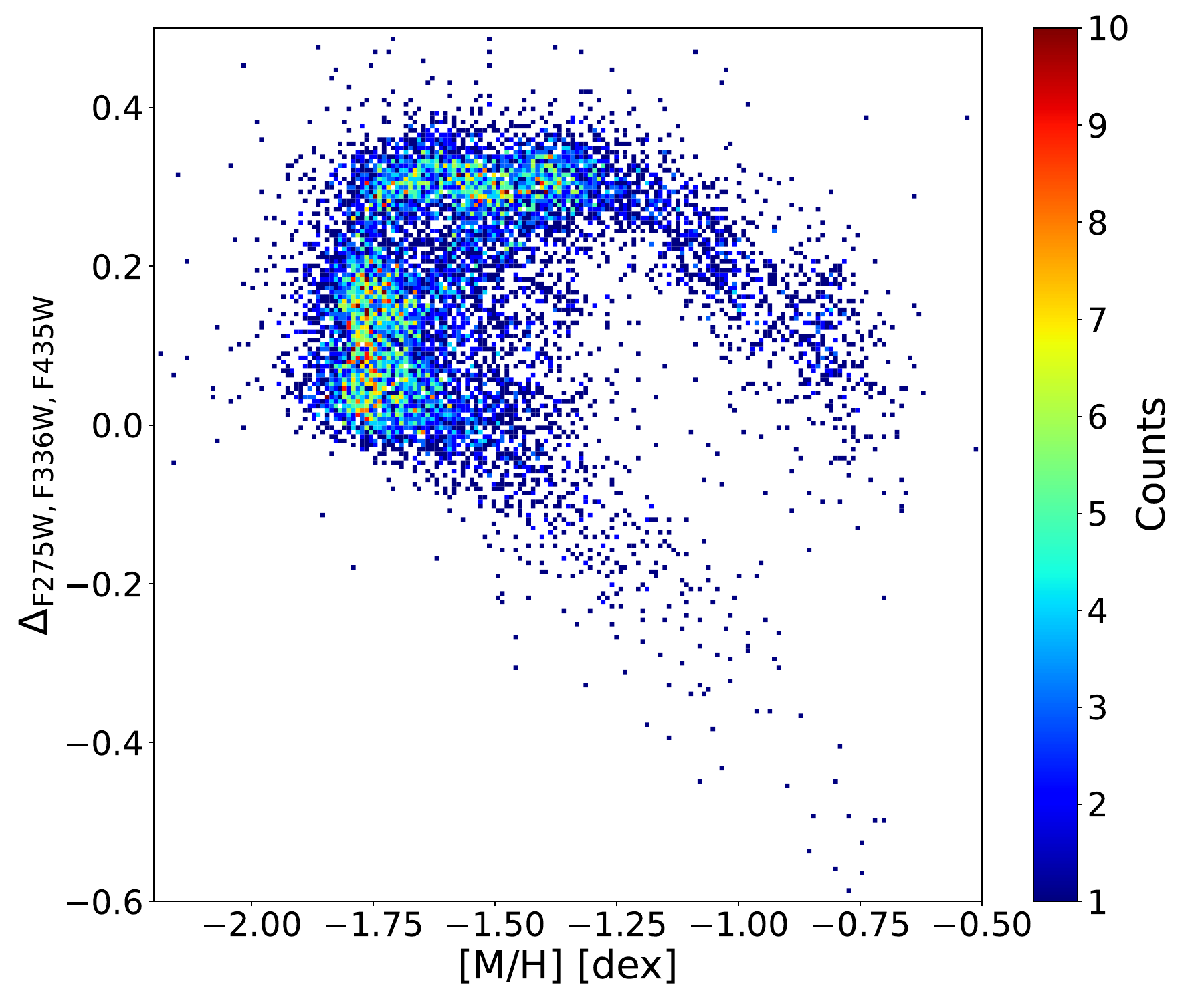}
\caption{\textbf{$\Delta$ values as a function of [M/H]}. The chromosome $\Delta$ values, left metallicity against $\Delta_{\rm F275W,F814W}$ and right $\Delta_{\rm F275W,F336W, F435W}$ plotted against metallicity in a density histogram. On the left, the red line separates the two streams visible and the black line indicates the metal-rich stars that are excluded from the analysis in \autoref{sec:2seq}.\label{fig:vs MH}}
\end{figure*}

The x-axis of this chromosome map is mostly sensitive to stellar populations with different helium (He) content and metallicity (Fe), whereas the y-axis is mostly efficient in identifying stellar groups of stars with different nitrogen (N) abundance (see e.g. Figures 25 and 27 in \citealt{Marino_2019} and Figure 5 in \citealt{Milone_2020}, for details). In a typical GC, what is known as first (1G) and second-generation (2G) stars would lie around $\Delta_{\rm F275W,F814W} \sim 0$, with 1G being centered at the origin (0,0) and the 2G stars found above ($\Delta_{\rm F275W,F336W, F435W} > 0$). In $\omega$~Cen the most metal-poor stars follow this pattern, while higher metallicity stars have different tracks at $\Delta_{\rm F275W,F814W}>0$. Further abundance studies will help us distinguish the different sub-populations even better.

\subsubsection{Metallicity dependency}\label{sec:x,y vs MH}

Since we have the metallicity information for all these stars we can investigate the relationship between the $\Delta$-values and [M/H], see \autoref{fig:vs MH}. $\Delta_{\rm F275W,F336W, F435W}$ vs [M/H] shows as expected a similar structure to the chromosome map since $\Delta_{\rm F275W,F814W}$ tracks the [M/H] variations. Specifically [He/H] and [Fe/H] variations, are the main cause for the horizontal displacement in the chromosome map (see Figure 5 in \citealt{Milone_2020} and Figure 27 in \citealt{Marino_2019}).

Interestingly, we notice two distinct sequences or streams in the $\Delta_{\rm F275W,F814W}$ vs [M/H] diagram, which correspond to stars with different light-element abundances (see also \autoref{fig:2seq} in \autoref{sec:2seq}). Specifically, the upper stream is composed of the most extreme stars in chemical properties, enhanced in [Na/Fe], [Al/Fe] and depleted in [O/Fe], [C/Fe] with respect to the lower stream stars with the same metallicity \citep{Marino_2019, Milone_2020}. It is also known that the vertical displacement in the chromosome map is mainly caused by [N/Fe] \citep[e.g.,][]{Marino_2019, Milone_2020}, hence stars located in the upper stream have higher [N/Fe] abundance than stars in the lower stream. In addition, He-enhanced and Mg-depleted stars are expected in the upper stream \citep{Milone_2015b, Milone_2018, Marino_2019, Milone_2020}. Further, Fe enrichment occurs in both streams, the upper has a slightly higher peak \citep{Marino_2019}, which we will also show in \autoref{sec:1g vs 2g} and \autoref{sec:2seq}.

In the future we are aiming to perform a detailed study on the abundances of individual stars (Wang et al. in prep.), using DD-Payne \citep{Ting_2017, Xiang_2019} and further investigate the effect the abundance variations have on the chromosome map. That will allow us to identify the main cause for the two sequences seen in the left panel of \autoref{fig:vs MH}. We are also working on combining stellar spectra of the same subpopulations to identify the overall abundance differences between subpopulations (Di Stefano et al. in prep.), which will also help us tackle the cause of the two sequences.

\begin{figure*}[t!]
\plottwo{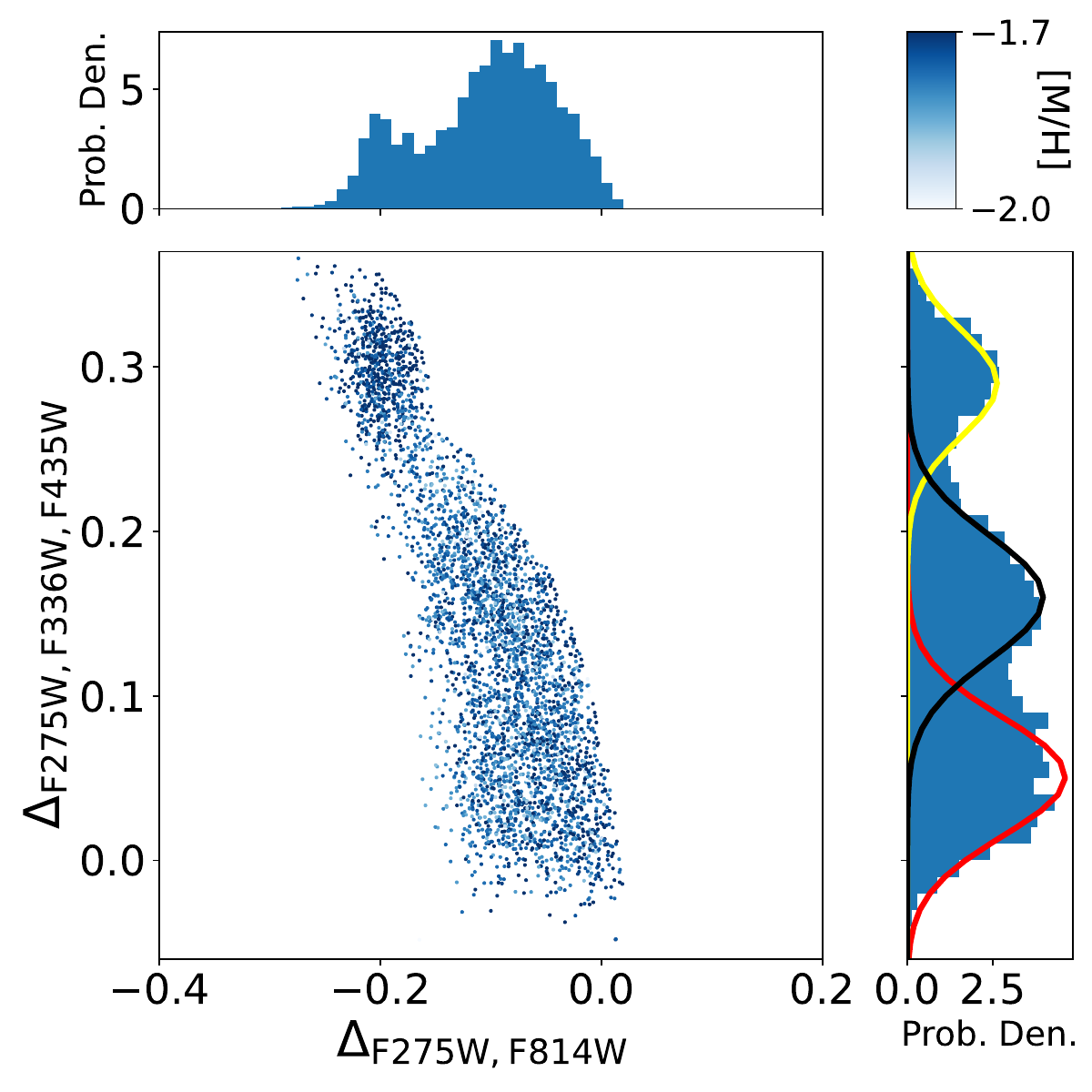}{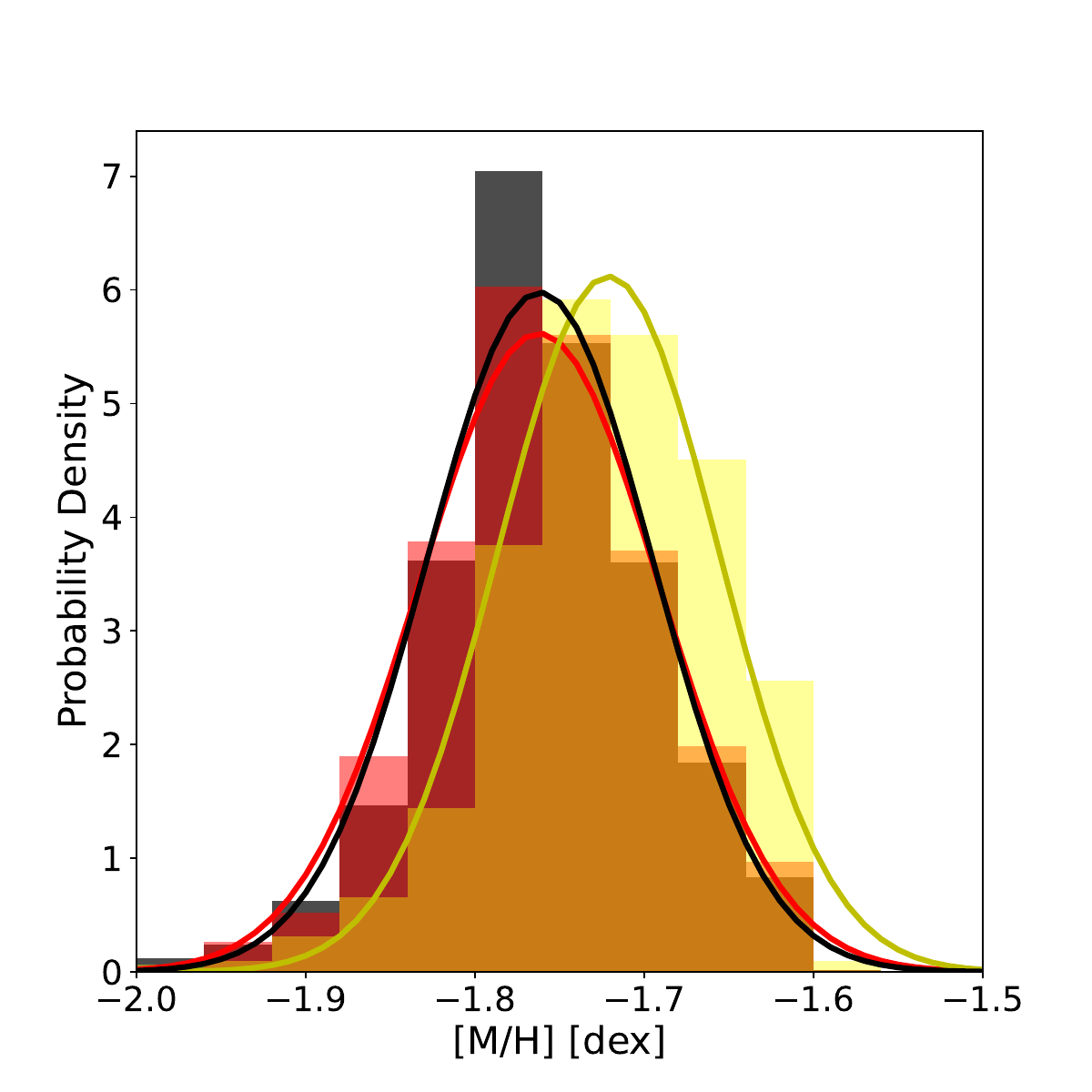}
\caption{\textbf{Multiple populations in the metal-poor stars.} On the left is the chromosome map only for the metal-poor stars with the probability density histograms of the $\Delta$ values. There we can identify 3 peaks in the $\Delta_{\rm F275W,F336W, F435W}$ histogram and plot their metallicity distribution on the right. The metallicity distributions for the populations indicated with the red and black lines are very close, while the population plotted in yellow exhibits a higher metallicity. \label{fig:gen}}
\end{figure*}

\begin{figure*}[t]
\gridline{\fig{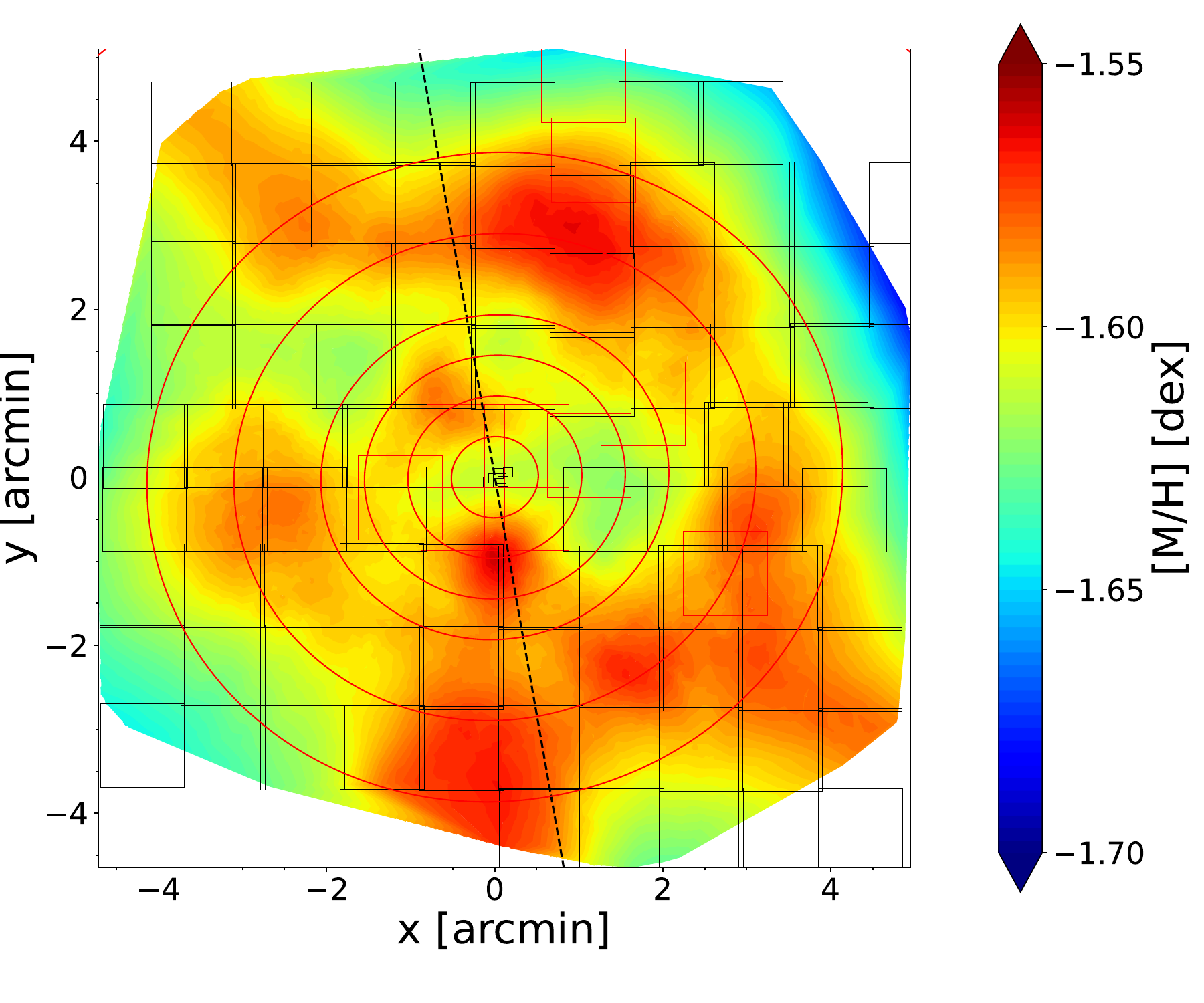}{0.5\textwidth}{(a.) LOESS map}}
\gridline{\fig{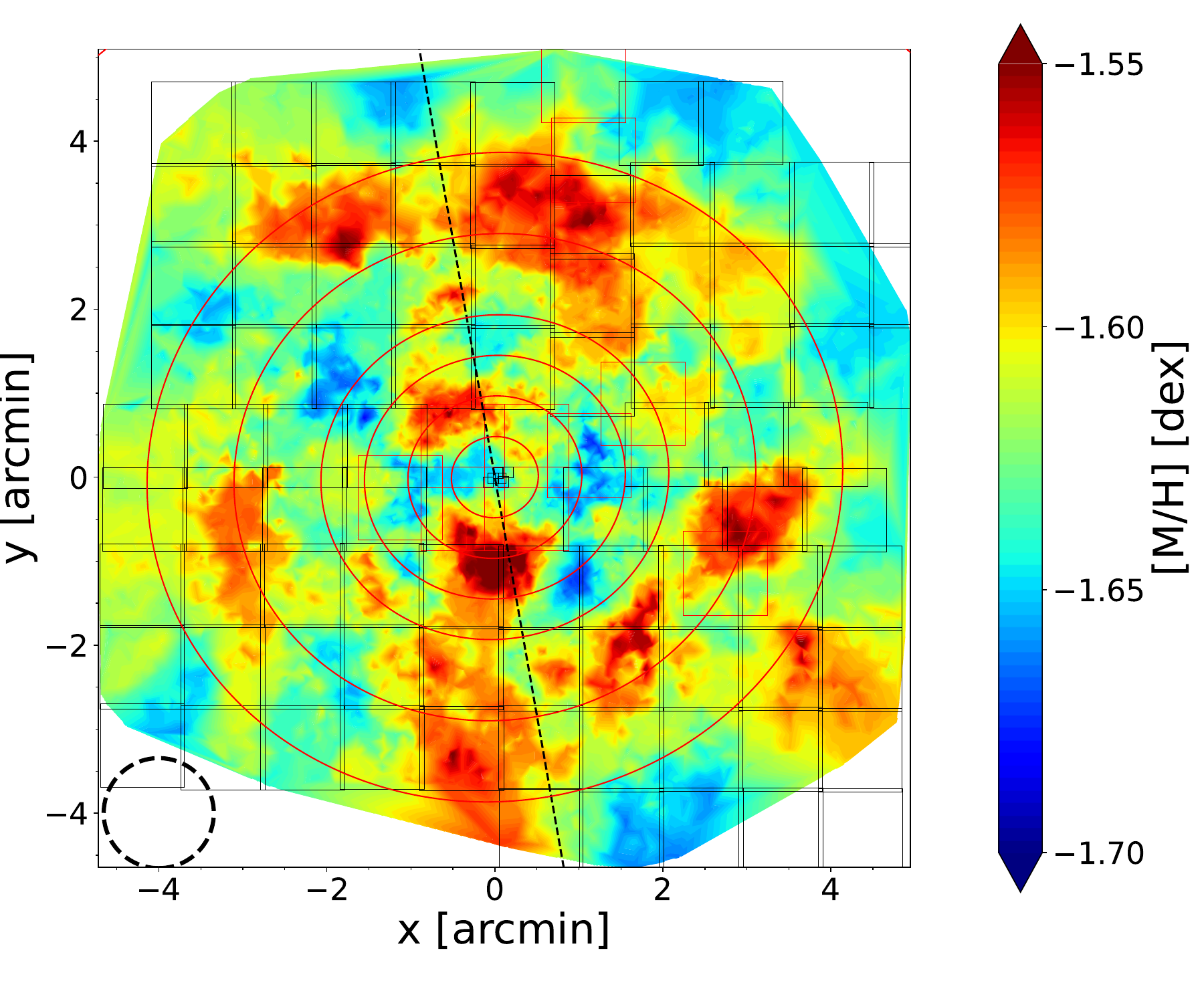}{0.5\textwidth}{(b.) Median map}
\fig{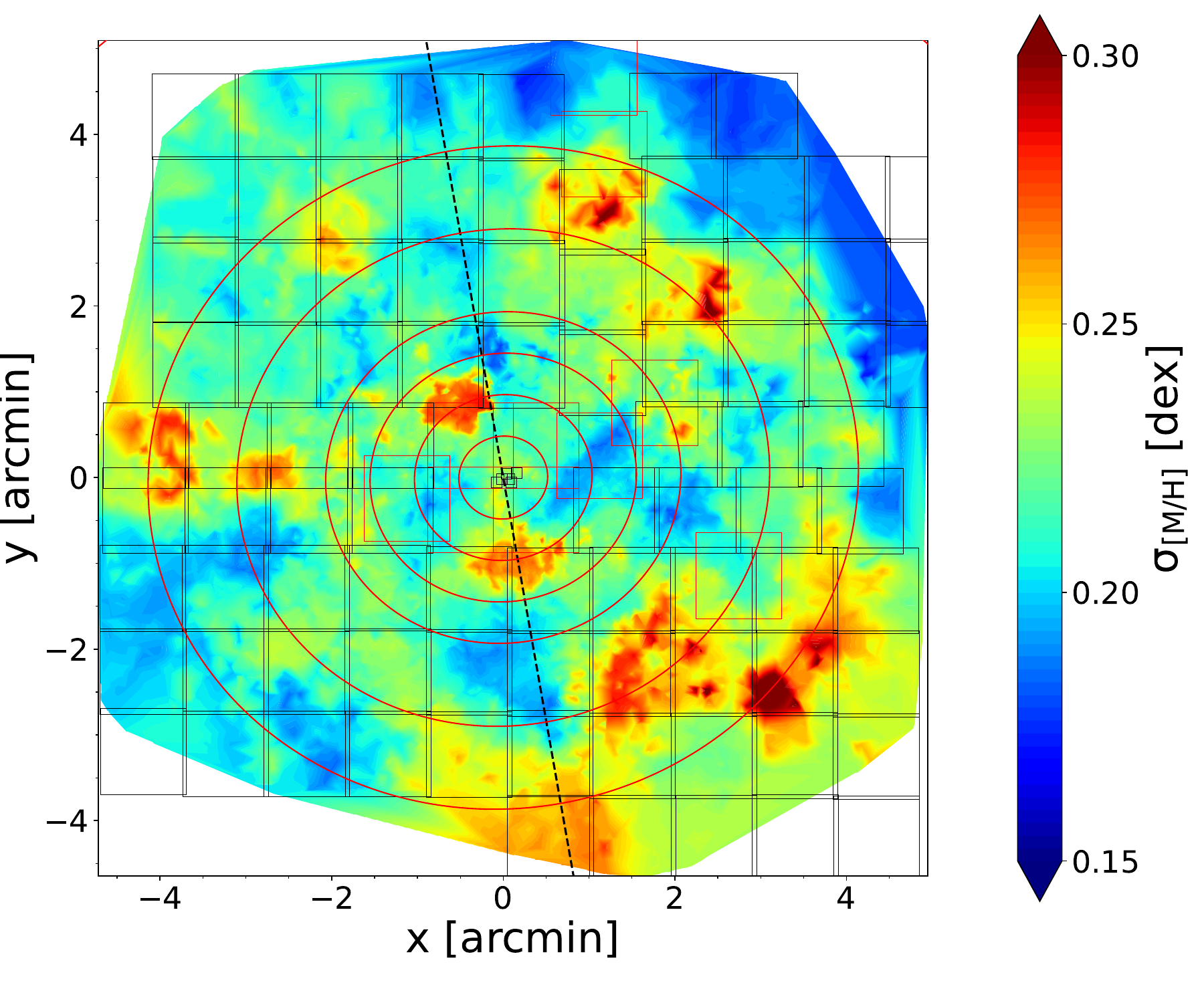}{0.5\textwidth}{(c.) Dispersion map}}
\caption{\textbf{Metallicity maps.} (a.) The LOESS smoothed metallicity map for visualization only, (b.) the median map from the 200 nearest neighbors, and (c.) the 68\% dispersion map from the 200 nearest neighbors. In (b.) the black dashed circle in the left corner shows the mean radius of the 200 neighbors. The pointing structure from MUSE is shown with faint squares, the red elliptical annuli are the bin edges used in \autoref{sec:grad}, and the minor axis of the elliptical annuli corresponding to the rotation axis is shown as a black dashed line.\label{fig:mh map}}
\end{figure*}

\subsubsection{The metal-poor component and its multiple populations}\label{sec:1g vs 2g}

The most metal-poor stars ([M/H] $  \lesssim -1.60$ dex) of the cluster follow a similar trend as the stars of mono-metallic globular clusters that separate in the so-called 1G and 2G stars along the $\Delta_{\rm F275W,F336W, F435W}$ axis due to light-element variations.

To identify these metal-poor stars in the chromosome map, we used stars in the metal poorest components of the GMM  (Index 1, 2, 4, 7, 11). Each star is assigned to the component that has the highest contribution to the GMM at the star's metallicity. This creates sharp metallicity cutoffs for the GMM populations. Then we remove blue or red outliers in the chromosome map (see the left panel in \autoref{fig:gen}) for the remaining stars. 

\begin{figure*}[t!]
\plottwo{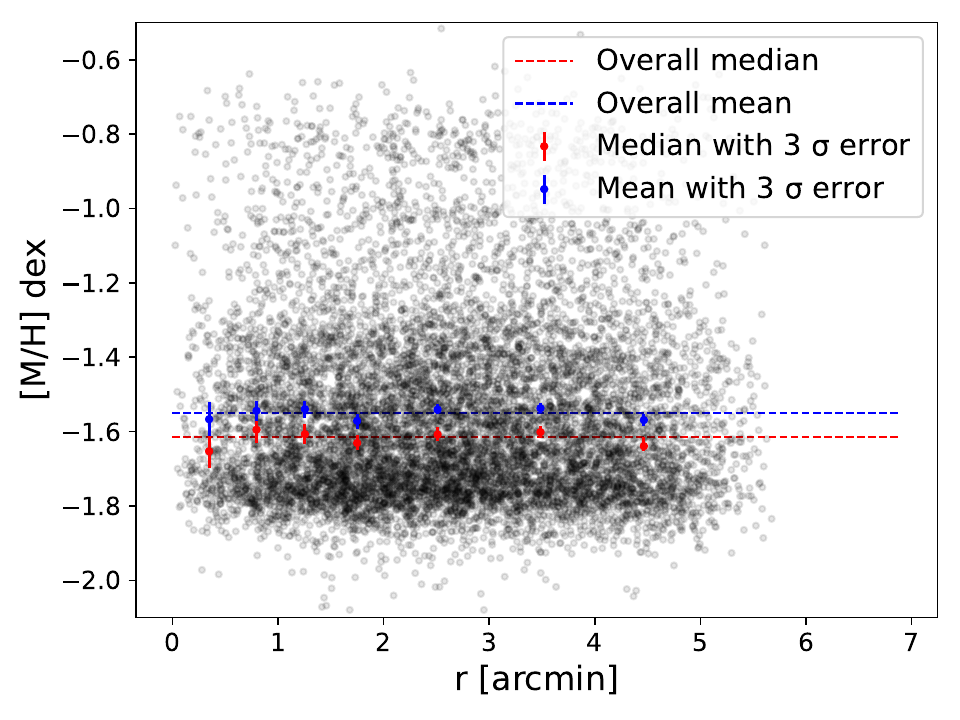}{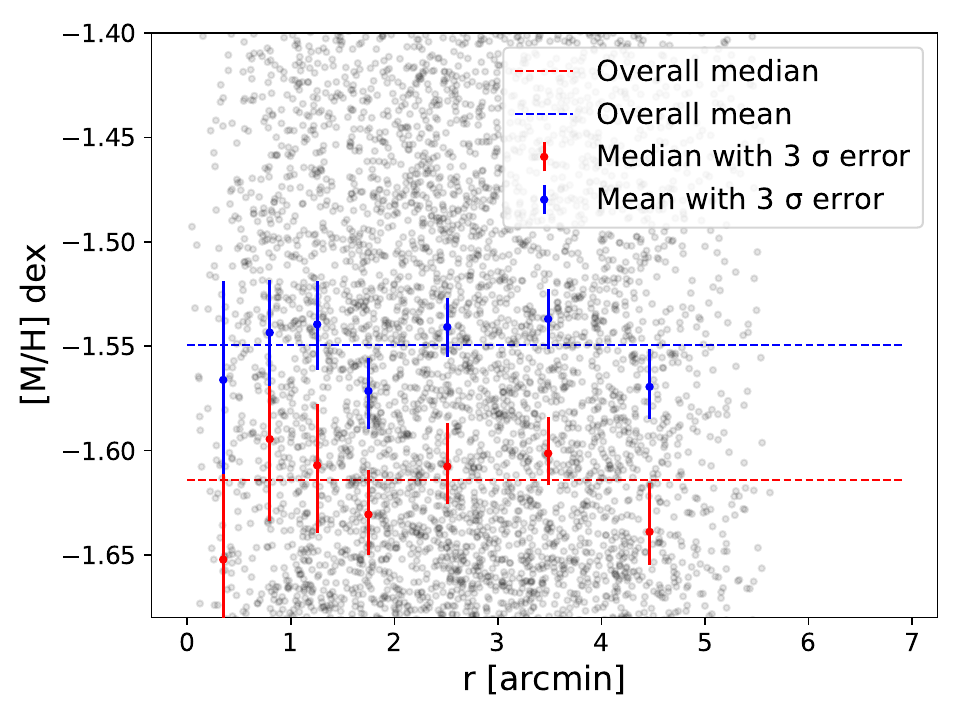}
\caption{\textbf{Overall metallicity gradient}. Black are all the stars used for this analysis red is the median value for each bin and blue is the mean value. The dashed lines show the total median or mean value for the whole sample. The left panel shows the whole [M/H] range while the right side is a zoom into metallicities around the mean and median.\label{fig: grad}}
\end{figure*}

If we only look at those stars, we can see that they also have a metallicity spread reaching from $-1.82$ to $ -1.68$~dex for 68\% of the stars, see \autoref{fig:gen}. With a simple Gaussian Mixture Model fit we identify 3 peaks (using BIC) in the $\Delta_{\rm F275W,F336W, F435W}$ distribution, separating them into three different populations. We see that the third peak (highest $\Delta_{\rm F275W,F336W, F435W}$ values) has the highest metallicity \citep[similar to the findings in][]{Marino_2019} and that all have a significant width. We investigate that further by looking at their metallicity distribution. The yellow peak has the highest mean metallicity at $ -1.721$~dex and an intrinsic standard deviation of $\sigma$ = 0.065~dex, the red is at $ -1.762$~dex and $\sigma$ = 0.071~dex and the black one is at $ -1.762$~dex and $\sigma$ = 0.067~dex. They have similar standard deviations [M/H] but the yellow peak is offset to higher [M/H], with a p-value below 1\% for the Anderson-Darling and Kolmogorov-Smirnov tests. The red and black distributions have higher p-values $>$25\% meaning that they are consistent to be drawn from the same distribution. Overall, even in this one metallicity component, there is a spread in [M/H] and it rises to higher [M/H] for higher $\Delta_{\rm F275W,F336W, F435W}$ values as seen also in \citet{Marino_2019}. However, since the difference is small, even though it seems to be significant (see also \autoref{fig:sub}), we do not have pure iron abundance but [M/H] which could be increased with enrichment in Na and Al which is expected for the black and even stronger for the yellow peak.

Higher $\Delta_{\rm F275W,F336W, F435W}$ also mean more enhanced Na and Al element abundances and depleted in O and Mg, as explained in \autoref{sec:x,y vs MH} and hence more He-rich stars \citep{Milone_2015b, Milone_2018, Marino_2019}. Therefore we will call stars belonging to the red peak Na-poor metal-poor (Na-poor MP) population, the yellow peak Na-rich metal-poor (Na-rich MP) population, and in the black peak are stars with intermediate Na enrichment.

There have been studies identifying sub-populations in $\omega$~Cen on the RGB \citep[e.g][]{Lee_1999, Pancino_2000, Sollima_2005b, Husser_2020} and the main sequence (MS) \citep[e.g.][]{Bedin2004, Bellini_2010, Milone_2017, Bellini_2017b}. However, connecting the RGB and main-sequence stars into the same subpopulations is a complex task that has not been done so far since the subpopulations overlap and cross each other below the turnoff point in the CMD. This will be the focus of future studies (Clontz et al. in prep).

\subsection{Spatial 2D Metallicity distribution}\label{sec: 2d}

We analyze the 2D spatial distribution of the metallicity since 2D variations might average each other out in radial bins.

We create smooth maps that still show details using the median of the 200 nearest neighbors (see \autoref{fig:mh map}) and for visualization only the locally weighted regression (LOESS) technique \citep[with a regularization factor f=0.1, ][]{Cappellari_2013}. We choose 200 nearest neighbors since that gives a smooth map without patches also in the velocity \citep[similar to][]{Pechetti_2024}. In \autoref{fig:mh map} we can see that indeed there is some structure, almost ring-like, consistently visible in both the LOESS and median map. In addition, we also plot half the difference of 68\% of the 200 nearest neighbors as an indication of the dispersion of the [M/H] values. Regions with lower median metallicity in the outskirts of the field have a relatively narrow metallicity distribution, while some of the structures with higher median metallicity also have a higher dispersion. However, there are also regions with moderately high median [M/H] and a narrow [M/H] distribution. In general, the variation in the dispersion map does not perfectly match the ring-like structure of the median metallicity map. Further, the pattern does not follow the MUSE pointing structure, reassuring us that the structure is not caused by calibration differences. We also verified that the structure does not appear if we take the ratio between the redmost RGB branch (metal-rich stars) and the blue RGB (metal-poor stars), which suggests that the structure is not caused by biases in our observations or selection of the sample in specific regions. However, this also indicates that the structure is not caused by the most metal-rich population but by some other intermediate [M/H] populations that might have been accreted later in its evolution and not yet well mixed.

\subsection{Metallicity gradients} \label{sec:grad}

\subsubsection{Overall cluster gradient}\label{sec:over grad}
Since the cluster is elongated in the plane of the sky, we do not use circular radial bins for further analysis but elliptical bins. We use $100^{\circ}$ for the positional angle (PA) \citep{VanDeVen_2006, Kamann_2018}, 0.07 for the ellipticity \citep[median value from][inside 5 arcmin]{Geyer_1983, Pancino_2003, Calamida_2020} and the equivalent radii for the bins are (0.5, 1, 1.5, 2, 3, 4, 7)~arcmin. These bin edges are overplotted in red in \autoref{fig:mh map}. In all the following analyses we calculate the median circular radius in each of the elliptical bins and show that in the figures.

\begin{figure*}[t!]
\plotone{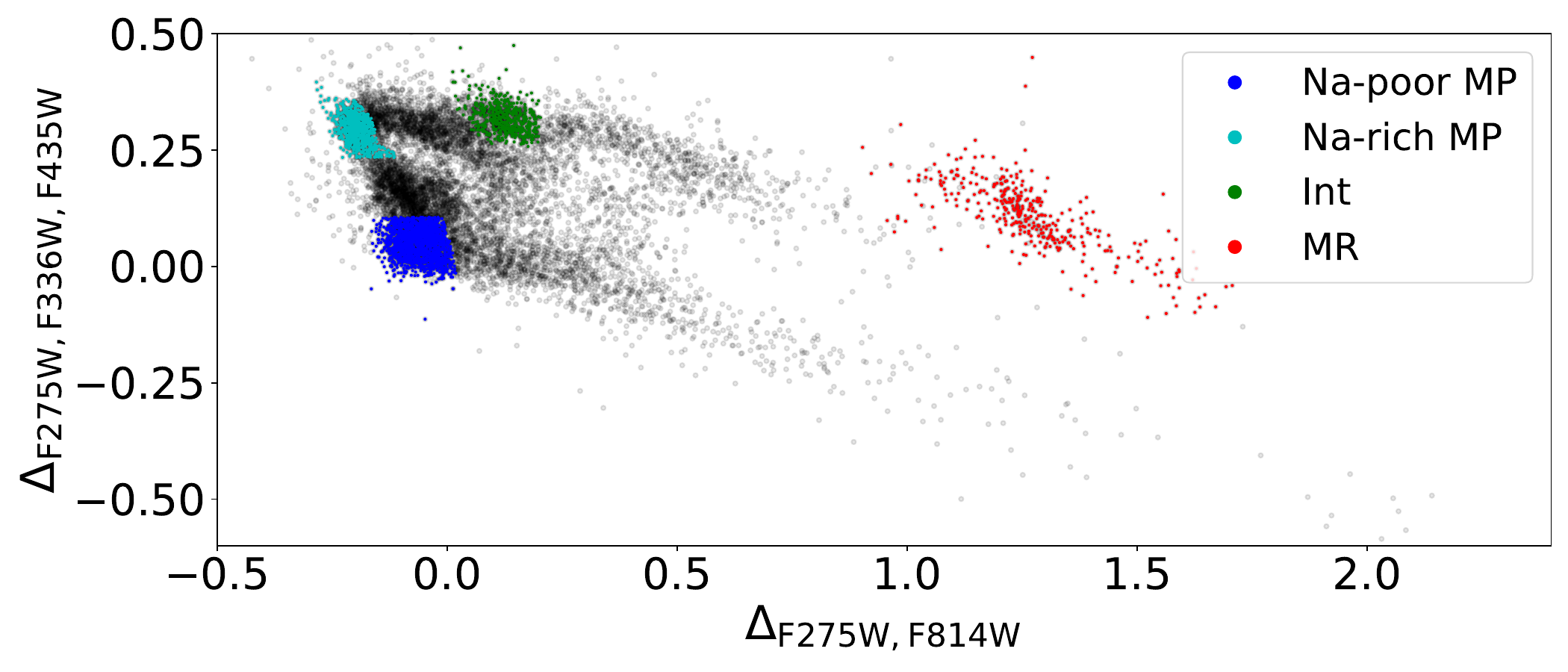}
\caption{\textbf{Chromosome map showing the four different subgroups considered in \autoref{sec:subpopulations}.} Red are the metal-rich stars, green the intermediate population, cyan the metal-poor but Na-rich, and blue the metal- and Na-poor stars. \label{fig:sub}}
\end{figure*}

\begin{figure*}[t]
\plottwo{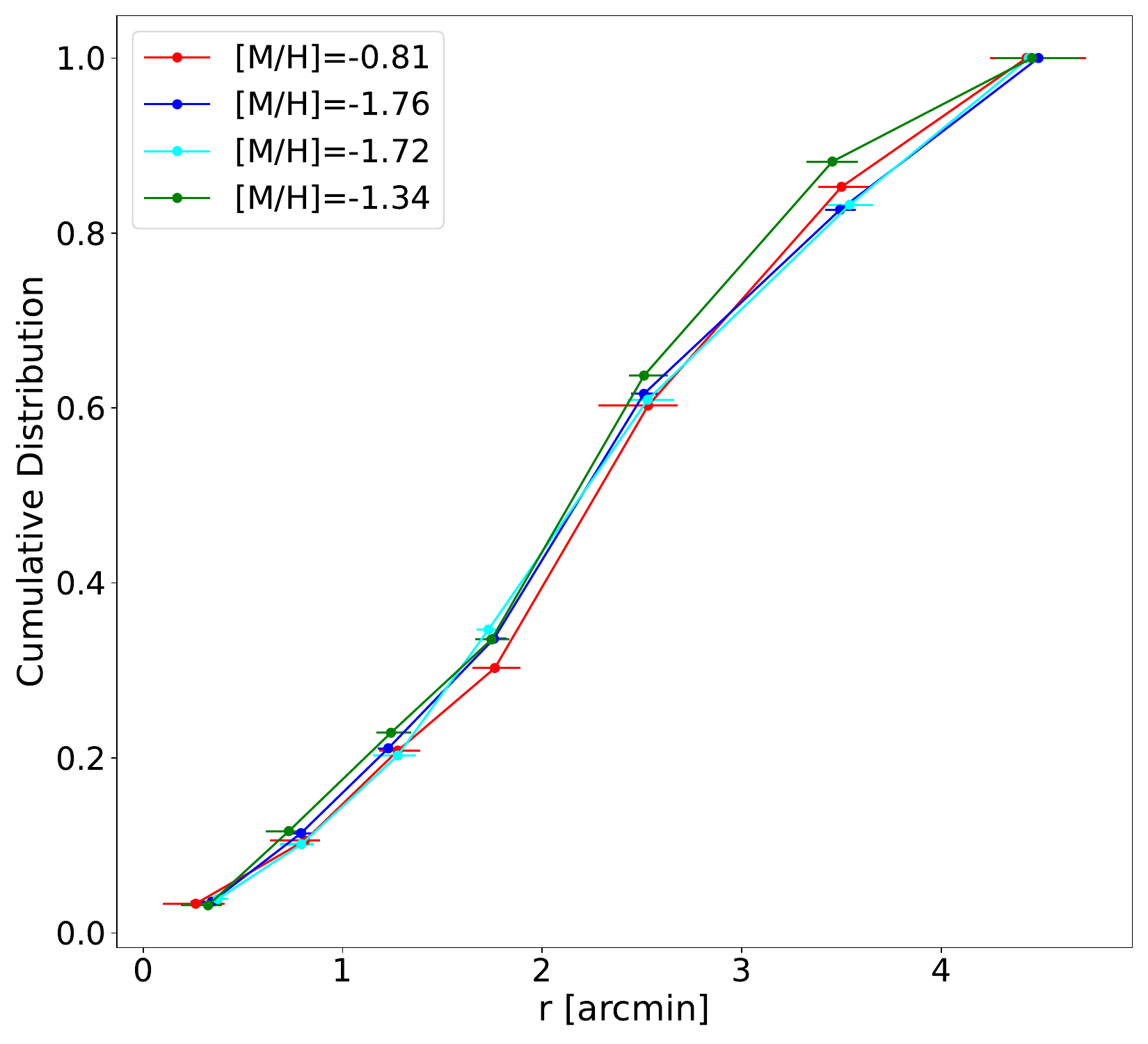}{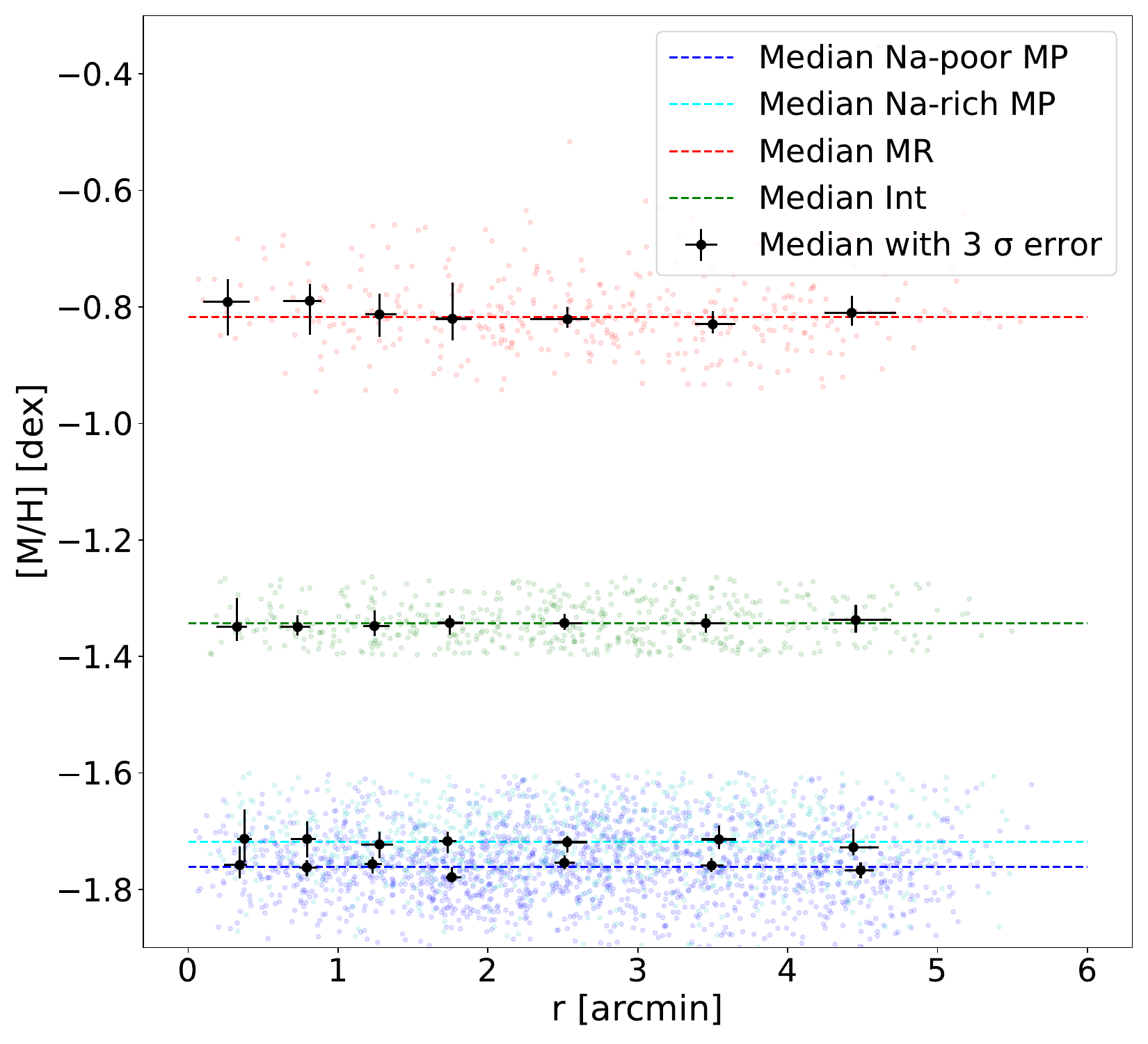}
\caption{\textbf{Spatial Distribution of the subgroups}. The left side shows the cumulative distribution of the four subgroups, the labels show the median [M/H] value. On the right is their metallicity versus radius.\label{fig:sub dist}}
\end{figure*}

In \autoref{fig: grad} we plot the mean and median metallicity value for each bin with the 99.7\% as 3$\sigma$ error bars, derived with bootstrapping. We can conclude that there is no significant gradient within the half-light radius of the cluster. However, there is an indication of the ring we see in the 2D distribution (\autoref{sec: 2d}), in the bins between 2-4~arcmin, where the median [M/H] is almost 3$\sigma$ (99.7\%) above the global median [M/H] of the cluster. Also, the last bin lies just outside the 3~$\sigma$ range below the overall median value. We see these same trends in the mean metallicity profile. The trend is not as strong as in the 2D maps since it is not a perfect ring at some lower and higher [M/H] areas cancel each other out when looking at the radial profile, but the signature is still visible. 

Previous studies \citep[e.g.][]{Sollima_2007, Bellini_2009, Calamida_2017, Calamida_2020} find spatial differences in sub-populations of the MS and the RGB. The bluer MS (associated with the intermediate metallicity RGB by \citet{Piotto_2005}), which is thought to be He-enriched \citep{Bedin2004, Norris_2004, Piotto_2005} is more centrally concentrated than the red MS (associated to the metal-poor RGB), but increasing towards the outer parts, above 25~arcminutes. While for the RGB the rich and intermediate follow the same distribution and are more centrally concentrated with a decline towards 8 to 10~arcminutes \citep{Bellini_2009}. The last bin in our metallicity gradient and lower metallicity towards the outer parts of the 2D map (see \autoref{sec: 2d}) indicate a similar trend with the more metal-rich or intermediate populations declining at the edge of the half-light radius. In \citet{Calamida_2017, Calamida_2020} they show that the more metal-poor stars follow the cluster elongation and are more numerous in the northern half, while the more metal-rich are elongated in the northeast-southwest direction and are more numerous in the eastern half. This trend is not visible in our data in the 2D distributions.

In general, for all these previous studies the spatial differences are prominent at the edge or after the half-light radius which is outside of our data range. From our analysis, inside the half-light radius, the sub-populations seem to be well mixed in radial bins.

\subsubsection{Spatial Differences between sub-populations}\label{sec:subpopulations}

Further, we want to investigate if there is a difference in the spatial distribution of different sub-populations. To this aim, we use four different subgroups, identified from the GMM and the chromosome map as follows. Na-rich but metal-poor (Na-rich MP), Na-poor, and metal-poor (Na-poor MP) populations were identified in \autoref{sec:1g vs 2g}, then we further select one intermediate metallicity (Int) population, index 7 in the GMM, and constrain it to the stars belonging to the clear overdensity in the chromosome map. We emphasize however that this is not the only intermediate metallicity population but one of many and we only selected this one because it is easily distinguishable in the upper sequence of the Chromosome map. Finally, the metal-rich population (MR) includes the most metal rich stars with a visible separate sequence in the chromosome map. See \autoref{fig:sub}  for their location on the chromosome map. 

This identification of four sub-populations is a first selection of some easily detectable sub-populations in the chromosome map that span a wide spread in [M/H] and are far apart in the chromosome map. We do not aim to find all subpopulations but to select a few to check their distribution over the field of view. We plan to use all available information and machine learning techniques to classify the subpopulations in an upcoming paper (Clontz et al. in prep.).

The cumulative radial distribution and metallicity gradient of each sub-population are shown in \autoref{fig:sub dist}. The Anderson-Darling test gives a p-value higher than 25\%, so the null hypothesis is not rejected and the sub-populations could have the same underlying distribution. The only exception is between the Na-poor MP or Na-rich MP and the Int populations, the p-values are lower ($\sim$ 8\%), however still too high to reject the null hypothesis. Future more detailed separation between sub-populations might reveal stronger differences. The populations do not show any significant metallicity gradient, see \autoref{fig:sub dist} right panel. The right panel also confirms our finding in \autoref{sec:1g vs 2g} that the metallicity difference between Na-rich and Na-poor metal-poor populations is significant (outside the 3$\sigma$ range) for all radii, except for the first bin, where due to the low number of stars the error bars are larger. The sub-populations, depending on metallicity or Na-enrichment, we selected are all well-mixed.

\begin{figure}[h!]
\plotone{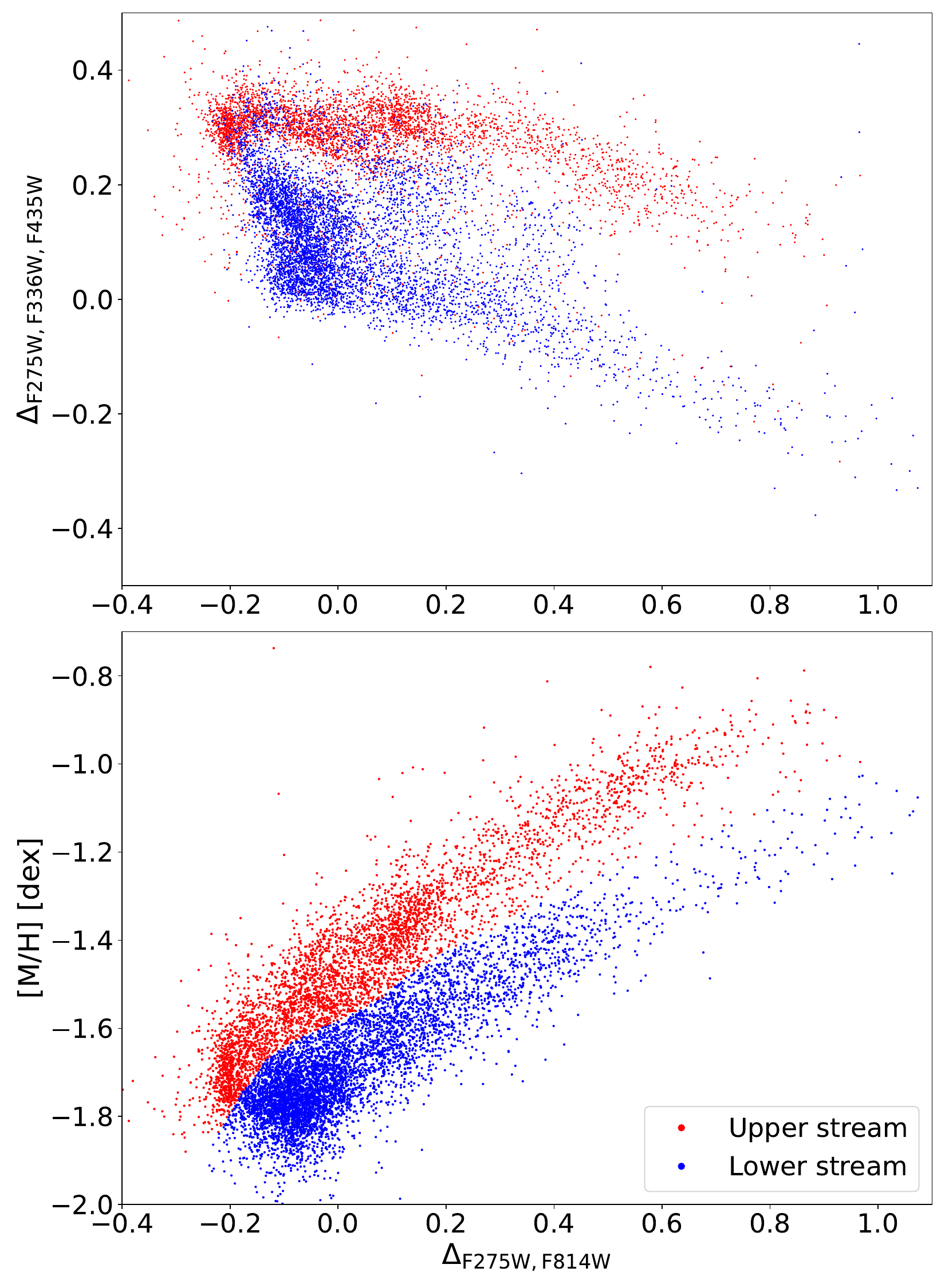}
\caption{\textbf{The two streams in [M/H] vs $\Delta_{\rm F275W,F814W}$.} The top panel shows the chromosome map and the bottom panel the $\Delta_{\rm F275W,F814W}$ vs [M/H]. We color-coded the stars depending on whether we identified them as belonging to the upper or lower stream in the original \autoref{fig:vs MH}. \label{fig:2seq}}
\end{figure}

\subsubsection{Differences in the two streams of [M/H] vs $\Delta_{\rm F275W,F814W}$}\label{sec:2seq}

\begin{figure*}[t]
\plottwo{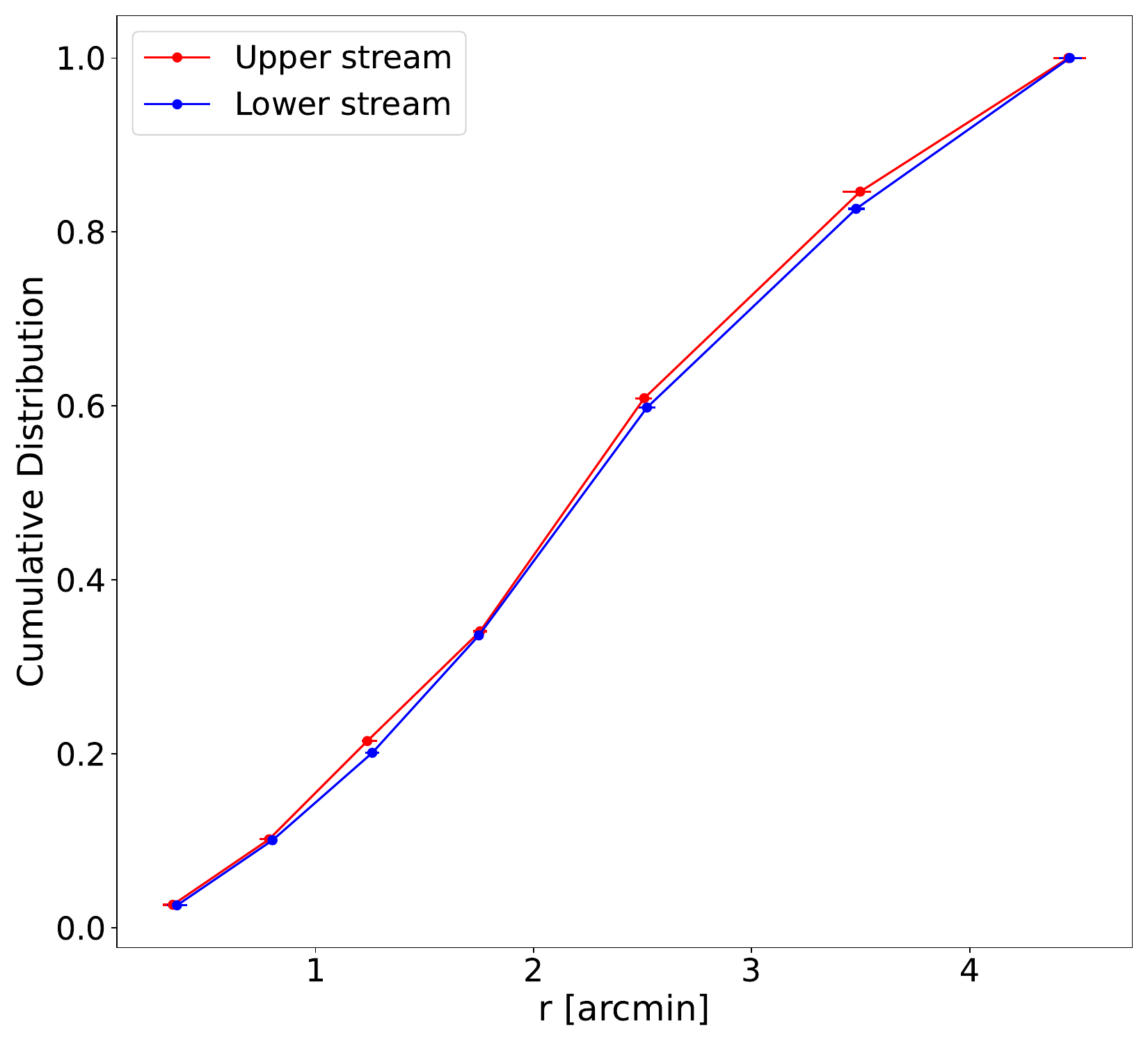}{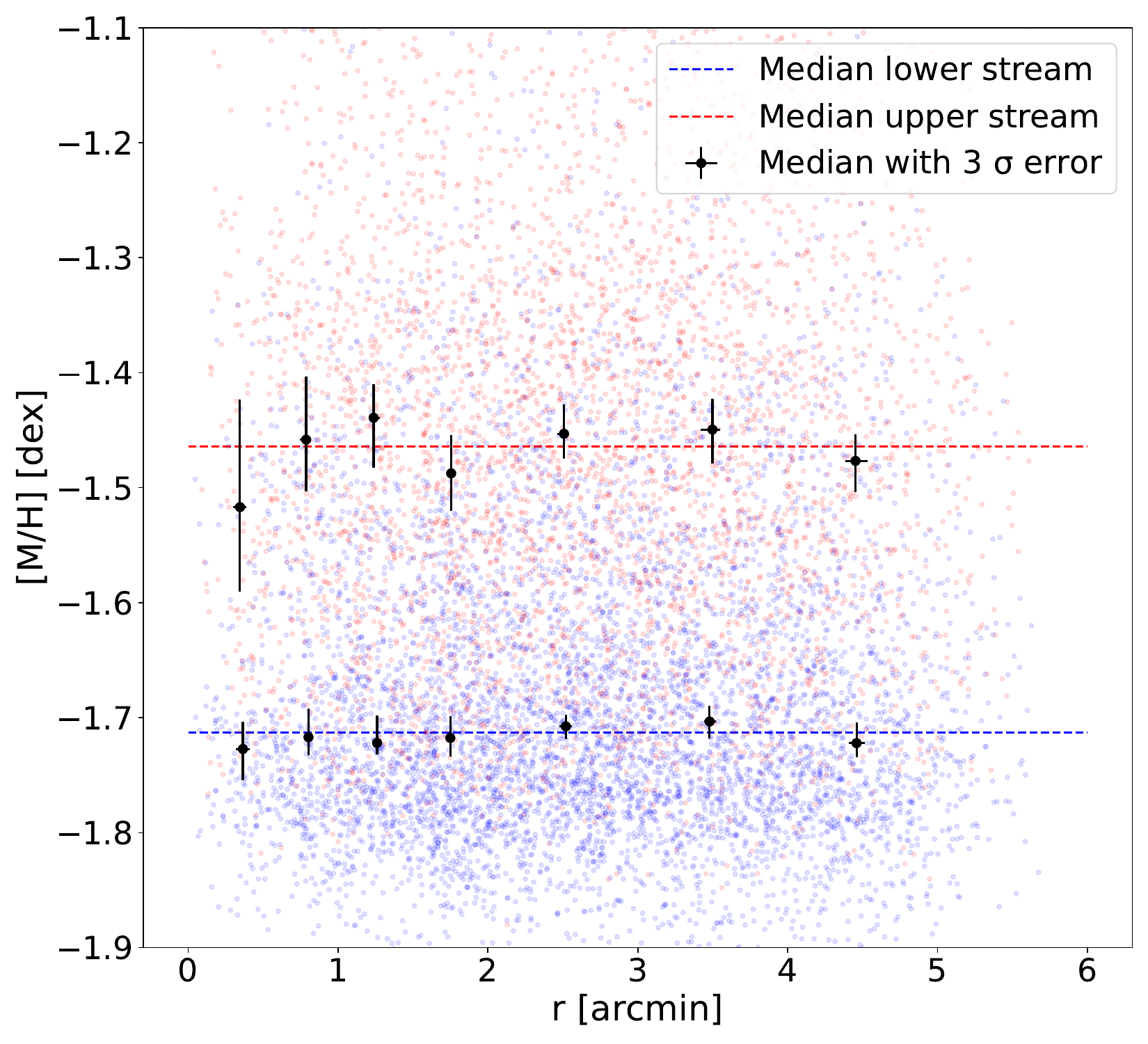}
\caption{\textbf{Spatial Distribution of the two streams in [M/H] vs $\Delta_{\rm F275W,F814W}$}. The left side shows the cumulative distribution of the two sequences. On the right is their metallicity versus radius.\label{fig:2seq dist}}
\end{figure*}

In the previous section, we chose four sub-populations with different metallicities and Na-enhancement and found no significant spatial difference. To further check the difference between sub-populations with different enhancement in light elements, we use the two streams seen in [M/H] vs $\Delta_{\rm F275W,F814W}$ (see \autoref{sec:x,y vs MH} and \autoref{fig:2seq}). We exclude the metal-rich stars (black line in \autoref{sec:x,y vs MH}) since they lie in between the sequences. The cumulative distribution is shown in the left panel of \autoref{fig:2seq dist} and they do not differ significantly, Kolmogorov-Smirnov and Anderson-Darling tests show 10\% and 8\% p-values confirming that the null hypothesis is not rejected for their spatial distribution. However, the p-values are lower than for the subpopulations indicating that there might be differences that could get stronger towards larger radii (similar to the difference of the blue and red  MS \citet{Sollima_2007, Bellini_2009}).

Their metallicity, however, is different with the upper sequence having a median metallicity higher than the lower sequence. That is not unexpected since most of the metal-richer stars lie on the upper stream and slightly higher iron values are expected in that stream as discussed in \autoref{sec:x,y vs MH}. However, again no significant metallicity gradient is visible (see right panel \autoref{fig:2seq dist}).

\section{Conclusion}\label{sec: concl}

We present a detailed study of the metallicity distribution of 11,050 RGB stars in $\omega$~Cen with a mean error of 0.041~dex and a median SNR$\sim$ 54. We find a mean value of $ -1.550$~dex and a median value of $ -1.614$~dex spanning 0.44~dex for 68\% of the stars with $m_{\rm F625W}<17$~mag. Further, we identify multiple peaks in the metallicity distribution indicating different metallicity subgroups. 

Additionally, we investigate the chromosome map and its dependencies with metallicity. We looked in detail at the most metal-poor group and identified three populations that show a similar spread in [M/H] $\sim$ 0.07~dex while the [M/H] increases for higher $\Delta_{\rm F275W,F336W, F435W}$.

We also studied the 2D spatial distribution of the [M/H] values and created metallicity maps showing a ring-like structure with higher values, while inside and outside the [M/H] values drop. Finally, we check for metallicity gradients and spatial differences in radial bins for different subgroups of stars. There we find no gradient in [M/H] over our field of view.

Overall, the different [M/H] populations seem to be well mixed and no strong gradient is visible in their radial profiles. However, more information on the abundance might help better separate them and further investigate their differences. In addition, the different sub-populations may be well mixed spatially, but their kinematics may be used to separate them, as it takes longer to erase the stars' memory of their original orbits.

We plan to determine elemental abundances (Di Stefano et al., in prep; Wang et al. in prep), and identify sub-populations (Clontz et al., in prep.) and ages (Clontz et al., in prep). This will bring us closer to uncovering the formation history of $\omega$~Cen, the nearest nuclear star cluster.

\section*{Acknowledgment}
We thank the anonymous referee for the helpful and constructive comments. The work was based on observations collected at the European Southern Observatory under ESO program 105.20CG.001. Further, based on archival and new observations with the NASA/ESA Hubble Space Telescope, obtained at the Space Telescope Science Institute, which is operated by AURA, Inc., under NASA contract NAS 5-26555. M.A.C. acknowledges the support from FONDECYT Postdoctorado project No. 3230727. SK acknowledges funding from UKRI in the form of a Future Leaders Fellowship (grant no. MR/T022868/1). A.B. acknowledges support from HST grants GO-15857 and GO-16777.
M.L. acknowledges funding from the Deutsche Forschungsgemeinschaft (grant LA 4383/4-1). A.F.K. acknowledges funding from the Austrian Science Fund (FWF) [grant DOI 10.55776/ESP542].  

\facilities{VLT:Yepun (MUSE), \textit{HST} (ACS)}

\software{\textsc{Astropy} v5.2.1
\citep{astropy:2013, astropy:2018, astropy:2022}, \textsc{Matplotlib} v3.7.1 \citep{Hunter:2007, Caswel2023}, \textsc{Pandas} v1.5.3 \citep{mckinney-proc-scipy-2010, reback2020pandas}, \textsc{Numpy} v1.20.3 \citep{harris2020array}, \textsc{SciPy} v1.10.1 \citep{2020SciPy-NMeth}, \textsc{Scikit-learn} v1.2.2 \citep{scikit-learn}}

\appendix

\begin{figure}[b]
\plottwo{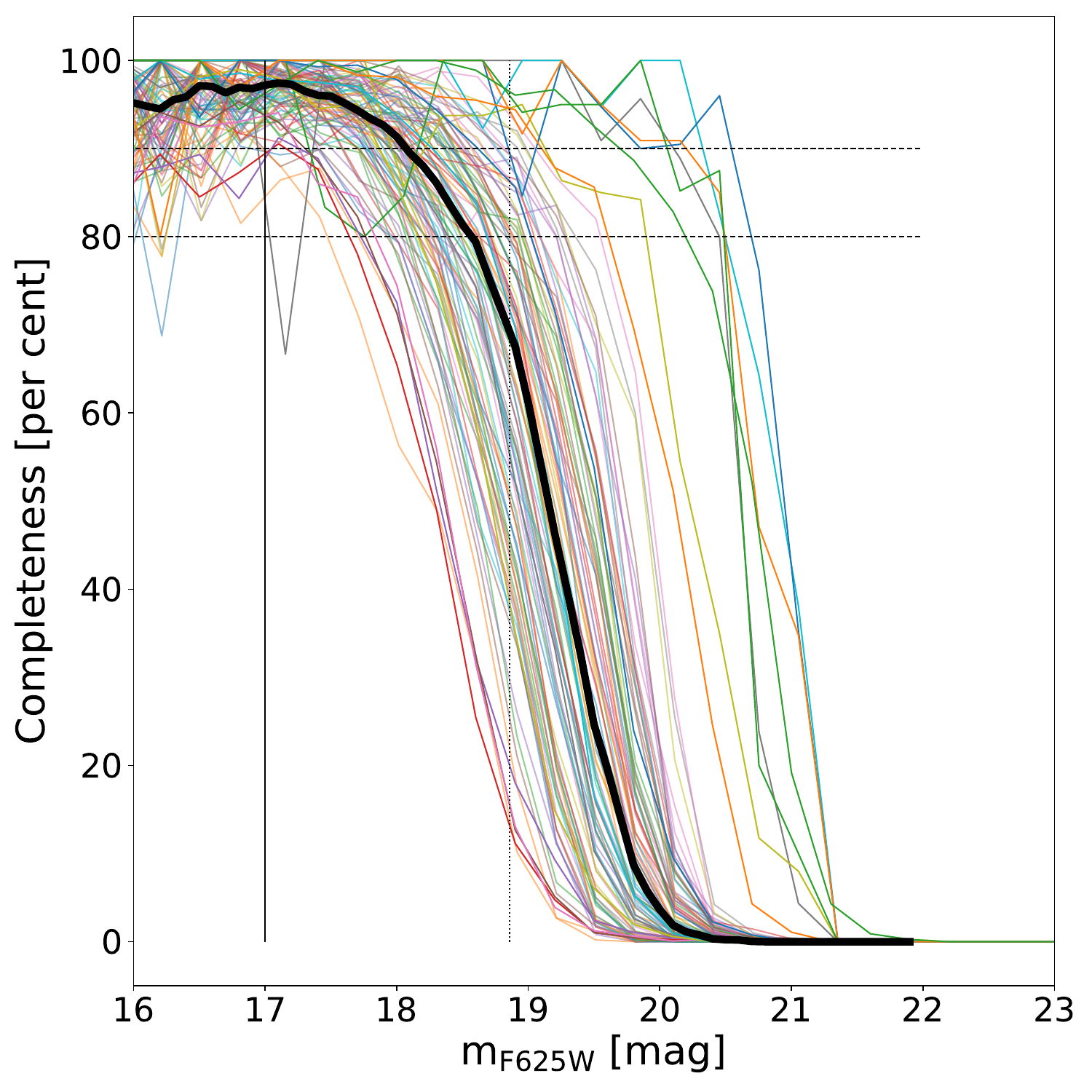}{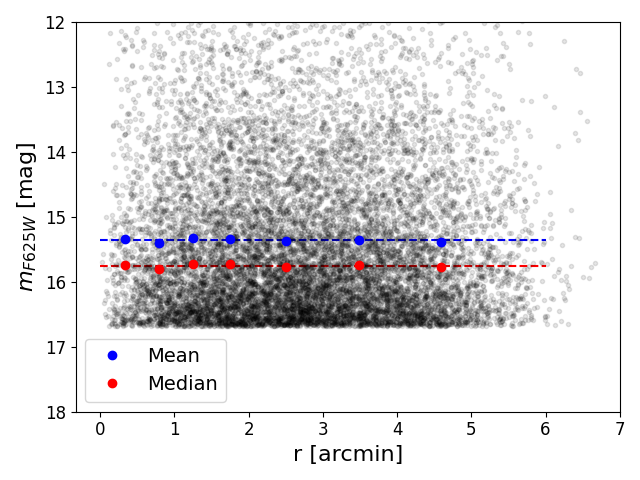}
\caption{\textbf{Completeness and magnitude}. Left plot: Each line shows the completeness fraction of one pointing compared to the \textit{HST} catalog from \citet{Anderson2010} and the thick black line shows the median completeness fraction. The vertical solid line is at 17~mag and the dotted one at 18.86~mag (for GO 50\% completeness on average). The dashed horizontal lines are at 80\% and 90\%. Right plot: the magnitude of the stars above 17~mag versus radius, which shows that for our quality cuts, there is no bias or gradient for specific magnitudes.\label{fig: compl}}
\end{figure}

\section{Completeness} \label{ap:compl}
In \citetalias{Nitschai_2023} we provided the completeness compared to the \citet{Anderson2010} catalog and found that at a magnitude of 18.86~mag in the F625W filter, we have 50\% completeness. For our study, we require higher completeness to avoid biases that could cause gradients or spatial variation in metallicity to vary. A magnitude cut of 17~mag gives us over 80\% completeness for all fields and even reaches 90\% for the majority, see \autoref{fig: compl} left plot. 

To further ensure that we do not bias our sample we check the radial dependency of the magnitude, which should be uniform. Indeed \autoref{fig: compl} (right) shows that the mean and median values remain constant over our whole field of view, not showing any crowding effects, for the RGB stars.

\section{[M/H] bias} \label{ap:bias}

\begin{figure}[t]
\gridline{\fig{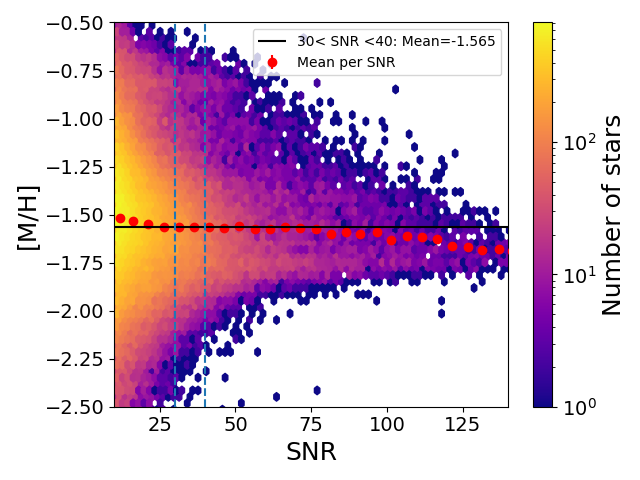}{0.4\textwidth}{(a.) GO}
          \fig{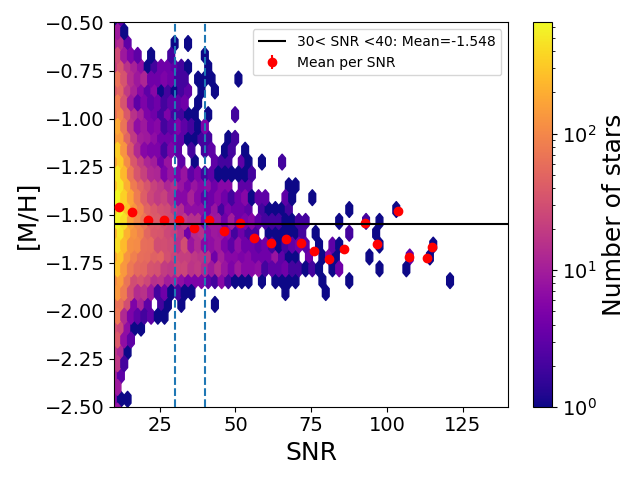}{0.4\textwidth}{(b.) GTO}}
\gridline{\fig{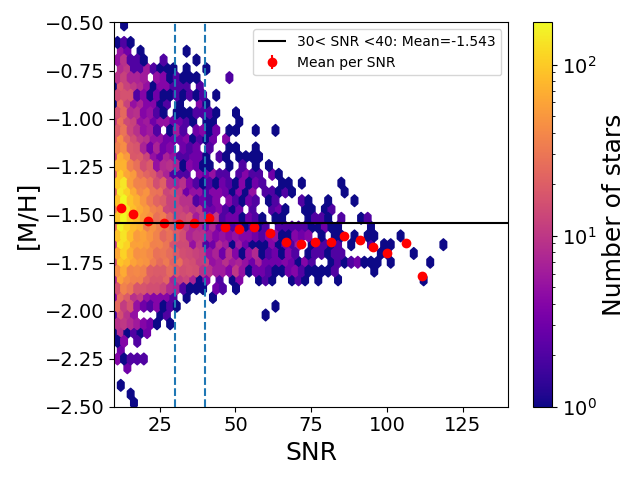}{0.4\textwidth}{(c.) Both, GTO\_GO}
          \fig{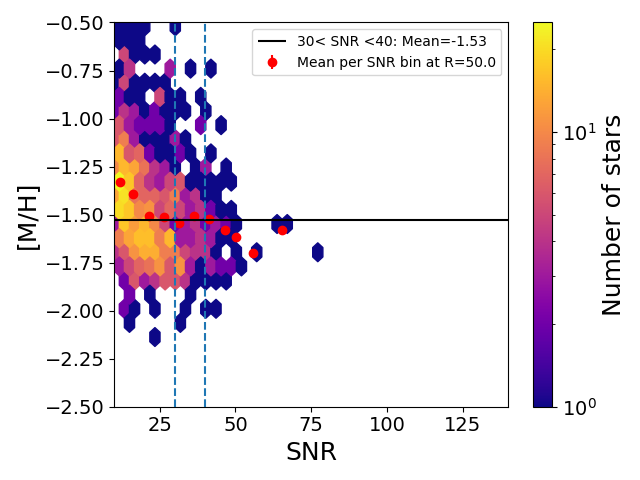}{0.4\textwidth}{(d.) NFM}}
\caption{\textbf{Metallicity bias.} We plot the metallicity versus the SNR for the different data sets underlying our spectroscopic catalog. Red dots are the mean values for each bin and the black solid line is the mean value between 30 and 40 SNR (indicated with blue dashed lines).\label{fig:MH bias data}}
\end{figure}

We noticed a metallicity bias in the different data sets, GO, GTO, NFM, when using all available magnitudes. In \autoref{fig:MH bias data} we can see that the [M/H] has always a trend to go to higher metallicities for lower SNR, but it differs between data sets. In general, the GTO stars with fainter magnitudes have higher [M/H] and lower SNR than the GO, which is likely caused by differences in the observing runs (e.g. different exposure times and number of exposures). However, the trend that stars with higher [M/H] have lower SNR is true for all stars and could be because they are redder, on average fainter (in general lower SNR), and have more absorption lines that make them look noisier to our spectral fitting routine compared to more metal-poor stars.
When excluding stars fainter than 17~mag in the F625W magnitude we remove most of the stars with SNR $\leq 20$ where the bias is visible, making our sample consistent between GTO and GO observations.

\begin{figure}[t!]
\centering
\includegraphics[scale=0.45]{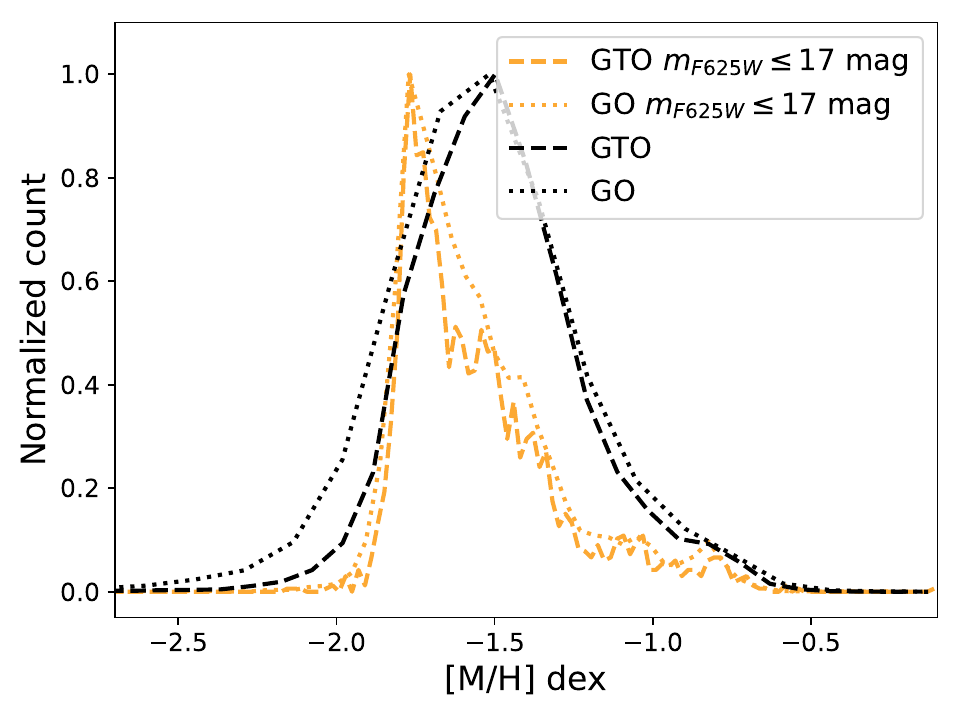}
\caption{\textbf{Metallicty distribution GO and GTO}. Using all magnitudes the two different data sets have important differences in their distributions. However, with a magnitude cut these disappear and the distributions look almost identical. \label{fig: MH GO GTO}}
\end{figure}

Additionally, we check if the distribution of the stars in GO and GTO are different, see \autoref{fig: MH GO GTO}. When using all magnitudes the GO distribution is broader reaching lower metallicities which would cause the GTO fields always to be on average more metal-rich. However, when applying the magnitude cut the distributions agree well not having any systematic difference.

\begin{figure*}[b]
\gridline{\fig{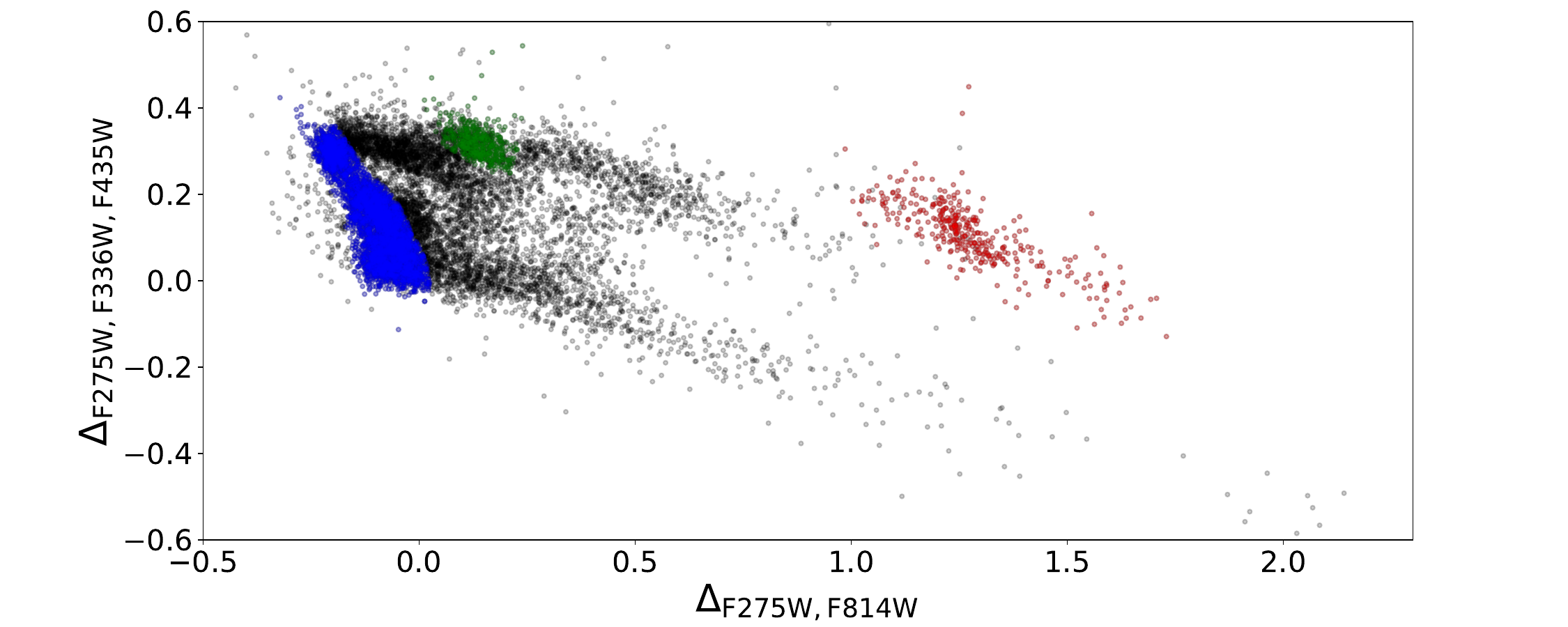}{0.5\textwidth}{a.} }
\gridline{\fig{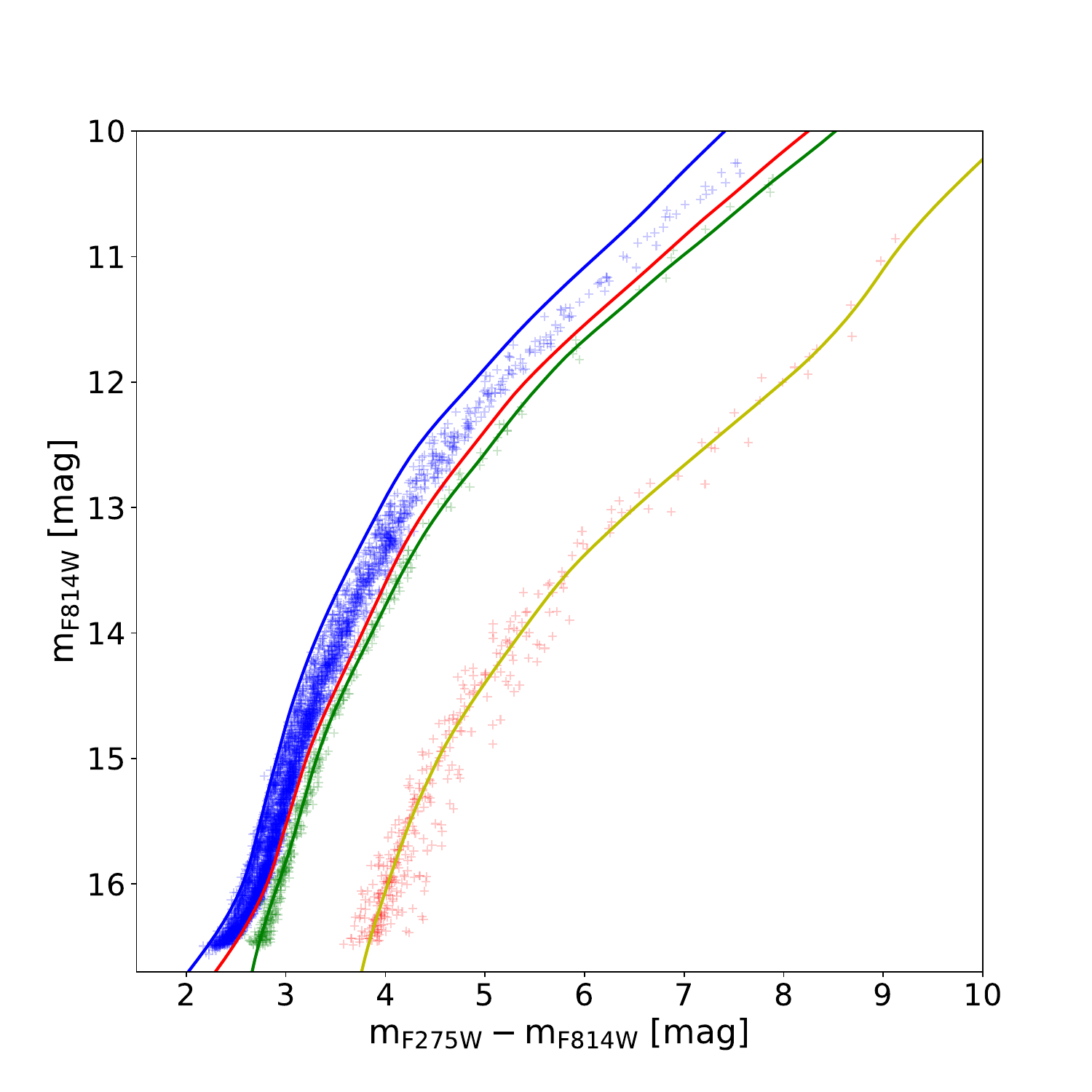}{0.4\textwidth}{b.}
          \fig{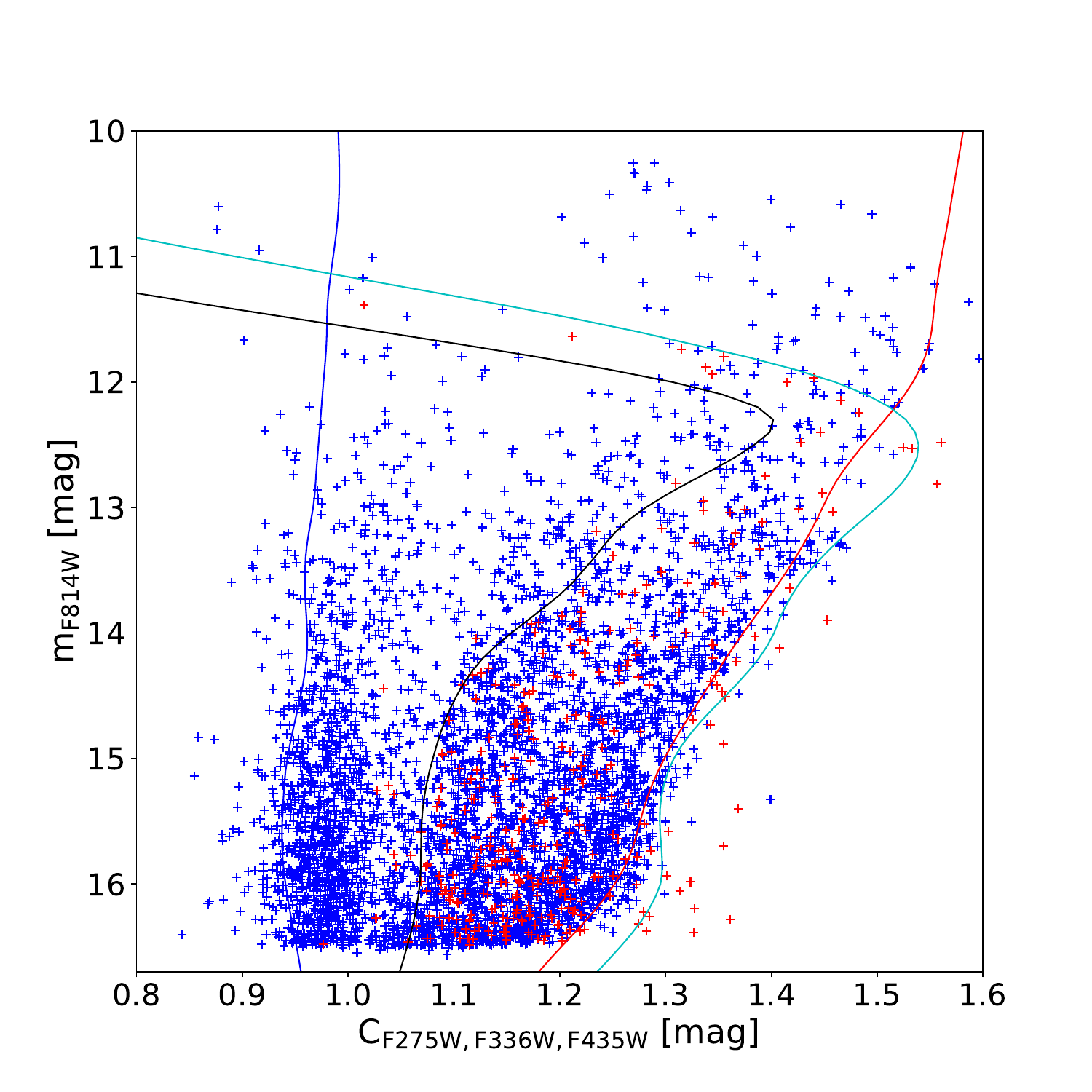}{0.4\textwidth}{c.}}
\caption{\textbf{The three subsamples used to calculate the chromosome map.} a. The chromosome map of $\omega$~Cen, color-coded with [M/H] only for the stars in the three subsamples. b. The CMD of the subsamples with the fiducial lines for each. Blue and red are the 96th and 4th percentile for the poor population (blue crosses), green is the median line for the intermediate (green crosses), and yellow is the median line for the rich population (red crosses). c. The pseudo-CMD of the subsamples with blue crosses being the poor and intermediate population and red the rich population. Blue and red lines are the 96\% and 4\% enclosing lines for the blue crosses and black and cyan for the red crosses.} \label{fig: sub fiducial}
\end{figure*}

\section{Chromosome map}\label{sec: detail chr} 

Since $\omega$ Cen is more complex than typical globular clusters, a more elaborate analysis is needed to create a useful chromosome map. Our method is adapted from the procedure described in \citet{Milone_2017}.

In summary, one needs more than just two, red and blue, fiducial lines since there are multiple different populations in $\omega$~Cen. Hence, one identifies three subsamples, the metal-poor population that also on its own would look like a chromosome map for typical clusters, an intermediate population, and a metal-rich population (a. \autoref{fig: sub fiducial}). The process is iterative and we start by identifying the subsamples in [M/H] first. Metal-poor stars are those belonging to peaks 1, 2, 4, 7 and 11 of the Gaussian Mixture model in \autoref{sec:distribution}, metal-rich stars belong to peak 9, and the intermediate population are stars in the 6 components. We then use slightly adapted fiducial lines from \citep{Milone_2017} to get an initial chromosome map. Next, we identify the subsamples using the different components identified for the [M/H] distribution and also the photometry as in \citet{Milone_2017}. Including the photometry information for the subsamples narrows CMD tracks and we calculate new fiducial lines. 

The red and blue fiducial lines correspond to the 96th and 4th percentiles in the CMD and pseudo-CMD.
In the $\rm m_{\rm F275W} -m_{\rm F814W}$ CMD (b. \autoref{fig: sub fiducial}) the red and blue fiducial lines correspond to the metal-poor population. For the intermediate and rich population, one needs to find the median fiducial line. In the CMD with the pseudo-color, $\rm C_{\rm F275W,F336W, F435W}$, the populations are more mixed and the intermediate and poor populations can use the same red and blue fiducial lines, while the metal-rich have different 96th and 4th percentiles indicated with the black and cyan lines. For all fiducial lines, we smoothed and corrected by hand for bright magnitudes.

\begin{figure*}[h!]
\plottwo{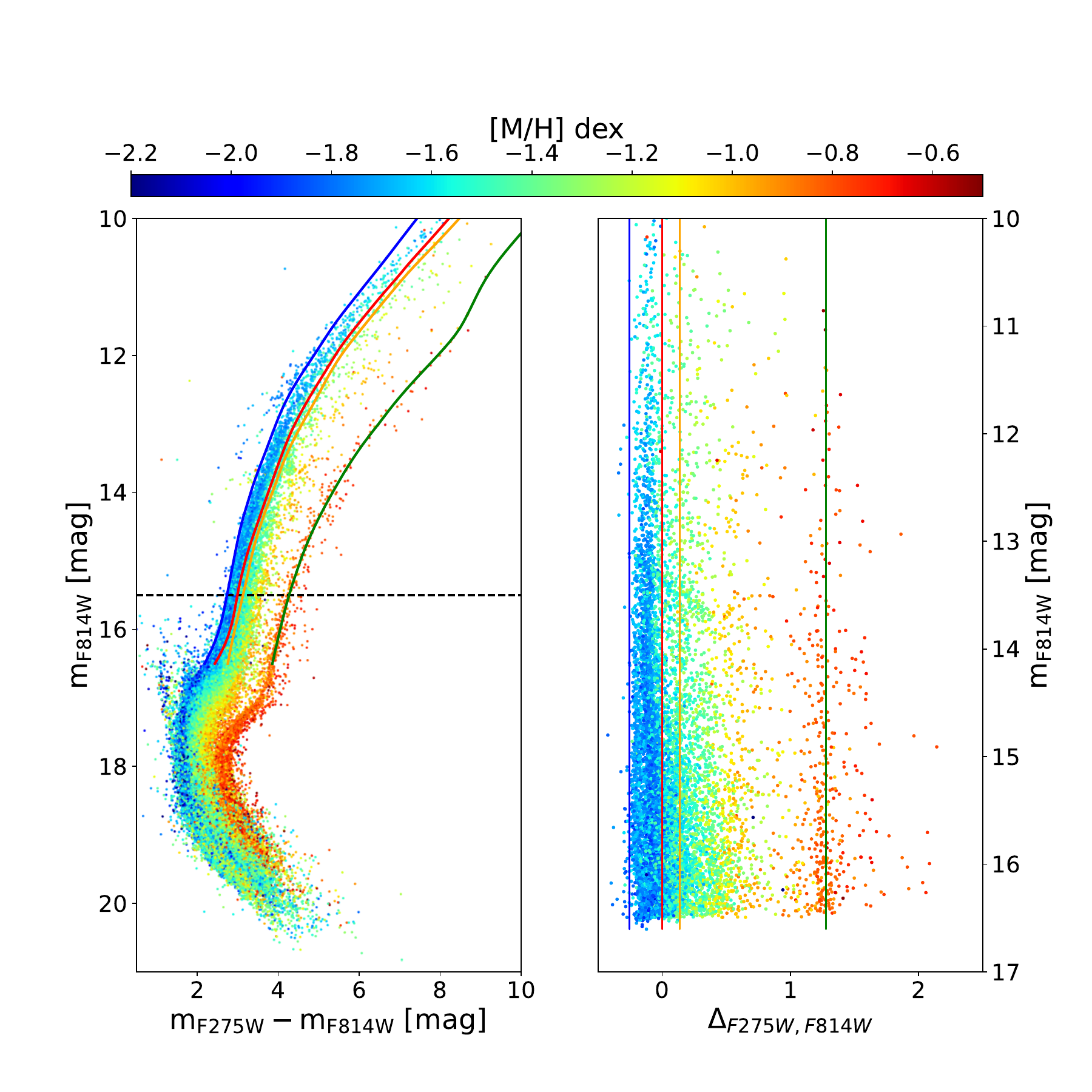}{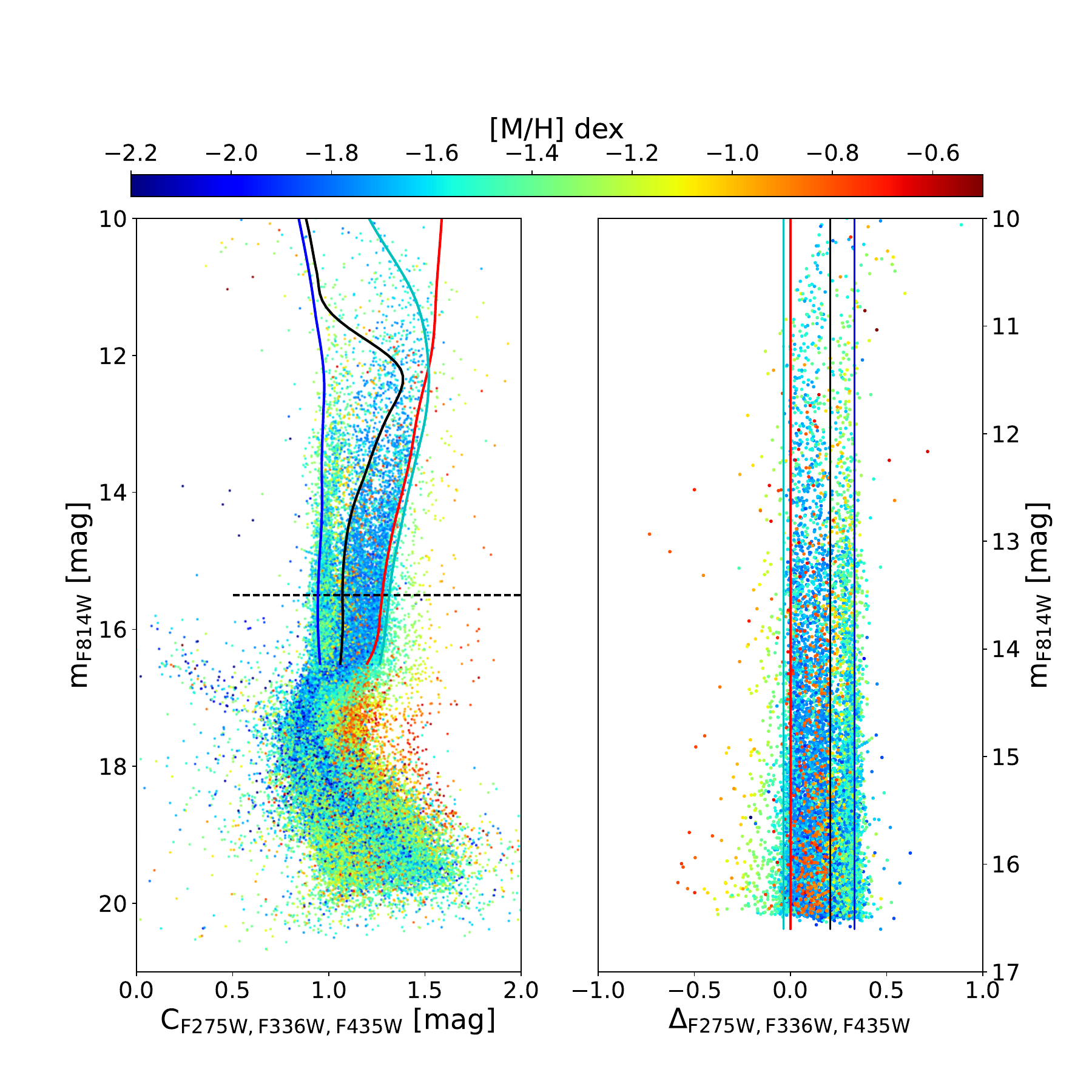}
    \caption{\textbf{Final Fiducial lines.} Left is the CMD for all stars and on the right is the pseudo-CMD both color-coded with [M/H] and showing the fiducial lines for the RGB. On the right of each subfigure is the verticalized CMD proving that the subsamples do follow these straight lines.}
    \label{fig:fiduacial_CMD}
\end{figure*}

We repeated this process twice refining our subsample selection and fiducial lines. Our final fiducial lines are shown in \autoref{fig:fiduacial_CMD} together with the verticalization of the colors for all bright stars.

To get the chromosome map values we calculate the following weights:
\begin{eqnarray}
    W_{\rm F275W, F814W_1} = X_{fiducialRed}(15.5~mag) -X_{fiducialBlue}(15.5~mag)\\
    W_{\rm F275W, F814W_2} = X_{fiducialOrange}(15.5~mag) -X_{fiducialRed}(15.5~mag)\\
    W_{\rm F275W, F814W_3} = X_{fiducialGreen}(15.5~mag) -X_{fiducialOrange}(15.5~mag)
\end{eqnarray}
 where e.g. $X_{fiducialRed}(15.5~mag)$ is the color of the fiducial line interpolated at 15.5 magnitude in F814W.  Then we can find the respective $\Delta_{\rm F275W,F814W_i}$ values:
\begin{eqnarray}
    \Delta_{\rm F275W,F814W_1} = W_{\rm F275W, F814W_1} \frac{X-X_{fiducialRed}}{X_{fiducialRed}-X_{fiducialBlue}}\\
    \Delta_{\rm F275W,F814W_2} = W_{\rm F275W, F814W_2} + W_{\rm F275W, F814W_2} \frac{X-X_{fiducialOrange}}{X_{fiducialOrange}-X_{fiducialRed}}\\
   \Delta_{\rm F275W,F814W_3} = W_{\rm F275W, F814W_2} + W_{\rm F275W, F814W_3} + W_{\rm F275W, F814W_3} \frac{X-X_{fiducialGreen}}{X_{fiducialGreen}-X_{fiducialOrange}}
\end{eqnarray}
Which of these three $\Delta_{\rm F275W,F814W_i}$ is used for each star depends on its relative position to the different fiducial lines, see \autoref{tab:delta} for details.
For the $\Delta_{\rm F275W,F336W, F435W}$ value we calculate:
\begin{eqnarray}
    W_{\rm F275W,F336W, F435W_1} = Y_{fiducialRed}(15.5~mag) -Y_{fiducialBlue}(15.5~mag)\\
    W_{\rm F275W,F336W, F435W_2} = Y_{fiducialRed}(15.5~mag) -Y_{fiducialCyan}(15.5~mag)\\
    W_{\rm F275W,F336W, F435W_3} = Y_{fiducialCyan}(15.5~mag) -Y_{fiducialBlack}(15.5~mag)
\end{eqnarray}
and then
\begin{eqnarray}
    \Delta_{\rm F275W,F336W, F435W_1} = W_{\rm F275W,F336W, F435W_1} \frac{Y_{fiducialRed}-Y}{Y_{fiducialRed}-Y_{fiducialBlue}}\\
    \Delta_{\rm F275W,F336W, F435W_2} = W_{\rm F275W,F336W, F435W_2} + W_{\rm F275W,F336W, F435W_3} \frac{Y_{fiducialCyan}-Y}{Y_{fiducialCyan}-Y_{fiducialBlack}}.
\end{eqnarray}
Depending on each metallicity each star gets assigned either $\Delta_{\rm F275W,F336W, F435W_1}$ or $\Delta_{\rm F275W,F336W, F435W_2}$. The condition is described in \autoref{tab:delta}.

\begin{deluxetable}{lcl}
\tablewidth{2.0\columnwidth}
\tablenum{3}
\tablecaption{Final chromosome map values \label{tab:delta}}
\tablehead{
\colhead{Final chromosome $\Delta$ value} & \colhead{$\Delta_i$ value} & \colhead{Condition}}
\startdata 
 &  $\Delta_{\rm F275W,F814W_1}$, if & $\Delta_{\rm F275W,F814W_1}\leq0$, \\
 & & right of the red fiducial line\\
$\Delta_{\rm F275W,F814W} =$ &  $\Delta_{\rm F275W,F814W_2}$, if & $\Delta_{\rm F275W,F814W_1}>0$ \& $\Delta_{\rm F275W,F814W_2}\leq W_{\rm F275W, F814W_2}$, \\
& & between red and orange fiducial line\\
&  $\Delta_{\rm F275W,F814W_3}$, if & $\Delta_{\rm F275W,F814W_3}> W_{\rm F275W, F814W_2}$, \\
& & right of the orange fiducial line\\
& & \\
$\Delta_{\rm F275W,F336W, F435W} =$ &  $\Delta_{\rm F275W,F336W, F435W_1}$, if & star does not belong to the most metal-rich  peak ($\neq$ index 9)\\
&  $\Delta_{\rm F275W,F336W, F435W_2}$, if & star belongs to metal-rich peak (= index 9)\\
\enddata
\tablecomments{The metal-rich peak (index 9) is taken from the Gaussian Mixture model \autoref{tab:multiGaus} in \autoref{sec:distribution}.}
\end{deluxetable}

\begin{figure}[h!]
\plotone{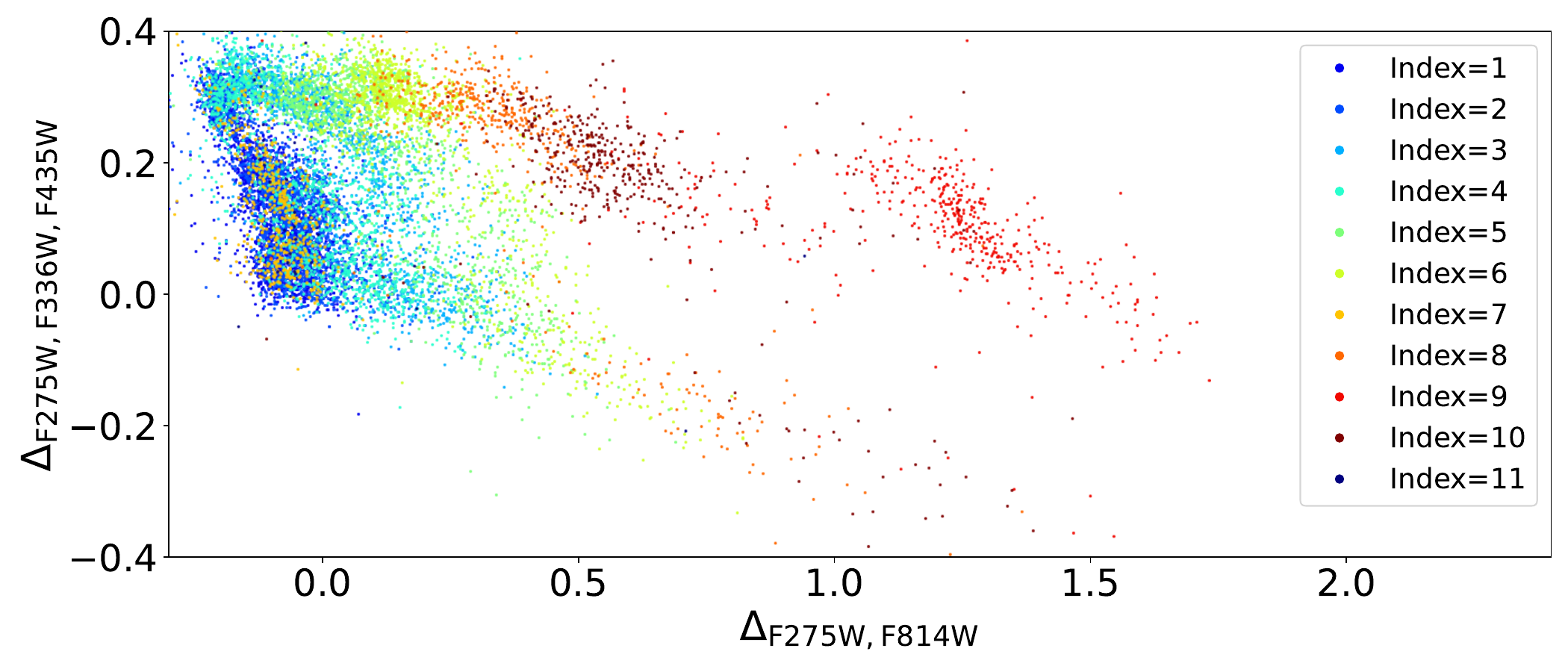}
\caption{\textbf{Chromosome map color-coded with the subgroups identified in the metallicity distribution.} We use the components of the Multi Gaussian Mixture Model used in \autoref{sec:distribution} and color-coded the stars in the chromosome map with the component (index) they most likely belong to.\label{fig: chrom mgm model}}
\end{figure}

Finally, we color-coded the final chromosome map with all the components of the Gaussian Mixture model of the [M/H] distribution (\autoref{sec:distribution}) in \autoref{fig: chrom mgm model}. All the components are at specific locations on the map depending on their [M/H], separating different sub-populations and showing the complexity of the clusters.

\newpage
\bibliography{paper}{}
\bibliographystyle{aasjournal}



\end{document}